\documentclass[twocolumn]{aastex631}
\newcommand{\teff}{\mbox{$T_{\rm eff}$}}
\newcommand{\logg}{\mbox{$\log g$}}
\newcommand{\msun}{\mbox{$\rm M_{\odot}$}}

\newcommand{\logl}{\mbox{$\textrm{log} (L/\textrm{L}_{\odot})$}}

\newcommand\bc[1]{\textcolor{black}{\textrm{#1}}}

\usepackage{hyperref}
\usepackage{amsmath}
\usepackage{makecell} 
\shorttitle{Revised extinctions and radii for 1.5 million stars}
\shortauthors{Yu et al.}
\graphicspath{{./}{figures/}}

\begin{document}

\title{Revised extinctions and radii for 1.5 million stars observed by APOGEE, GALAH, and RAVE}

\newcommand{\orcidauthorK}{} 
              
\author[0000-0002-0007-6211]{Jie Yu}
\affiliation{Max-Planck-Institut f{\"u}r Sonnensystemforschung, Justus-von-Liebig-Weg 3, 37077 G{\"o}ttingen, Germany}
\author[0000-0002-2604-4277]{Shourya Khanna}
\affiliation{Kapteyn Astronomical Institute, University of Groningen, Groningen, 9700 AV, The Netherlands}
\affiliation{INAF - Osservatorio Astrofisico di Torino, via Osservatorio 20, 10025 Pino Torinese (TO), Italy}
\author{Nathalie Theme\ss l}
\affiliation{Landessternwarte K{\"o}nigstuhl (LSW), Heidelberg University, K{\"o}nigstuhl 12, 69117 Heidelberg, Germany}
\author[0000-0002-1463-726X]{Saskia Hekker}
\affiliation{Landessternwarte K{\"o}nigstuhl (LSW), Heidelberg University, K{\"o}nigstuhl 12, 69117 Heidelberg, Germany}
\affiliation{Heidelberg Institute for Theoretical Studies (HITS) gGmbH, Schloss-Wolfsbrunnenweg 35, 69118 Heidelberg, Germany}
\author[0000-0002-0135-8720]{Guillaume Dr\'eau}
\affiliation{LESIA, Observatoire de Paris, PSL Research University, CNRS, Universit\'e Pierre et Marie Curie, Universit\'e Paris Diderot, 92195 Meudon, France}
\author[0000-0001-7696-8665]{Laurent Gizon}
\affiliation{Max-Planck-Institut f{\"u}r Sonnensystemforschung, Justus-von-Liebig-Weg 3, 37077 G{\"o}ttingen, Germany}
\affiliation{Institut f\"{u}r Astrophysik, Georg-August-Universit\"{a}t G\"{o}ttingen, Friedrich-Hund-Platz 1, 37077 G\"{o}ttingen, Germany}
\affiliation{Center for Space Science, NYUAD Institute, New York University Abu Dhabi, PO Box 129188, Abu Dhabi, UAE}
\author[0000-0002-7642-7583]{Shaolan Bi}
\affiliation{Department of Astronomy, Beijing Normal University, Beijing 100875, People's Republic of China}








\begin{abstract}
Asteroseismology has become widely accepted as a benchmark for accurate and precise fundamental stellar properties. It can therefore be used to validate and calibrate stellar parameters derived from other approaches. Meanwhile, one can leverage large-volume surveys in photometry, spectroscopy, and astrometry to infer stellar parameters over a wide range of evolutionary stages, independently of asteroseismology. Our pipeline, \texttt{SEDEX}\footnote{\url{https://github.com/Jieyu126/SEDEX}}, compares the spectral energy distribution predicted by the MARCS and BOSZ model spectra with 32 photometric bandpasses, combining data from 9 major, \mbox{large-volume} photometric surveys. We restrict the analysis to targets with available spectroscopy from the APOGEE, GALAH, and RAVE surveys to lift the temperature-extinction degeneracy. The cross-survey atmospheric parameter and uncertainty estimates are homogenized with artificial neural networks. Validation of our results with CHARA interferometry, HST CALSPEC spectrophotometry, and asteroseismology, shows that we achieve high precision and accuracy. We present a catalog of improved interstellar extinction ($\sigma_{A_V} \simeq$ 0.14 mag) and stellar radii ($\sigma_R/R \simeq$ 7.4\%) for  $\sim$1.5 million stars in the low- to high-extinction ($A_V \lesssim 6 $ mag) fields observed by the spectroscopic surveys.  We derive global extinctions for 184 Gaia DR2 open clusters, and confirm the differential extinction in NGC~6791 and NGC~6819 that have been subject to extensive asteroseismic analysis. Furthermore, we report 36,854 double-lined spectroscopic main-sequence binary candidates. This catalog will be valuable for providing constraints on detailed modelling of stars and for constructing 3D dust maps of the \textit{Kepler} field, the TESS CVZs, and the PLATO long duration observation fields.
\end{abstract}



\section{Introduction} \label{sec:intro}
Over the past decade, asteroseismology has significantly advanced several  fields of astrophysics. One of the main reasons is that asteroseismology can provide accurate and precise fundamental parameters, such as stellar radius, surface gravity, age, as well as distance and extinction \citep[see reviews by,][]{chaplin2013b, hekker2017a}.  However, this technique is limited as it requires long, \mbox{high-precision}, short-cadence light curves, which have so far been available only for a limited number of CoRoT \citep[$\sim$3,000,][]{de-assis-peralta2018a}, \textit{Kepler} \citep[$\sim$16,000,][]{yu2018a}, and \textit{K2} \citep[$\sim$19,000,][]{zinn2022a} stars. The ongoing TESS mission \citep{ricker2015a} has expanded the asteroseismic star sample by one order of magnitude, but is restricted to nearby stars \citep[$\sim$158,000, with a median distance of 800~pc,][]{hon2021a}. 

In the current era of \mbox{large-volume} photometric, spectroscopic, and astrometric surveys, one can exploit these complementary data sets to derive stellar parameters over various evolutionary stages, using a method independent of asteroseismology \citep[e.g.,][]{huber2017a, mints2017a, mints2018a, anders2022a}. One method to combine all these complementary data sets is to perform spectral energy distribution (SED) fitting, which is used in this work. This method involves matching the observed multiple broadband photometry with that predicted from model stellar spectra, to derive stellar parameters, such as bolometric flux and extinction. By combining these parameters with parallax, one can estimate stellar radius and luminosity. Recent examples of such implementation include studies for determining stellar parameters for Hipparcos and Tycho stars \citep{mcdonald2012b, mcdonald2017a}, subgiants in the TESS southern CVZ \citep{godoy-rivera2021a}, and dwarfs in general \citep{vines2022a}.

It is worth noting that SED fitting does not depend independently on effective temperature (\teff) and extinction; we refer to such variables as degenerate variables \citep[e.g.,][]{bailer-jones2011a, andrae2018a}. The studies mentioned above either assume zero extinction \citep{mcdonald2012b, mcdonald2017a} or use a typical extinction estimate for the entire sample \citep{godoy-rivera2021a}. These assumptions are legitimate for particular stellar samples, e.g. nearby stars, but have to be modified to study stars associated with a range of extinctions. One approach is to infer extinctions using \textit{a priori} known \teff, for example from spectroscopy with zero-point calibration if necessary. In the past decade, large spectroscopic surveys, such as APOGEE \citep{abdurrouf2022a}, GALAH \citep{buder2021a}, and RAVE \citep{steinmetz2020a}, have provided precise estimates of \teff, which in turn now allow us to derive precise extinction values.

The SED-fitting method offers unique advantages in deriving stellar parameters. First, it is independent of stellar evolutionary models. In comparison, traditional isochrone fitting methods rely heavily on evolutionary models that can introduce substantial systematics \citep{tayar2022a}. There have been persistent discrepancies between models and observations in terms of the determination of stellar parameters for late K- and M-type stars \citep{kraus2011a, feiden2012a, spada2013a, mann2015b, rabus2019a}. SED fitting has proven to be a robust method in this spectral-type regime, and has served to develop precise empirical relations to estimate radii and masses \citep{mann2015b}. Parameter degeneracy, however, prevents us from fully characterizing stars and any orbiting exoplanets, even with asteroseismology \citep{cunha2007a}. For example, helium fraction, mass, and radius are strongly coupled when using asteroseismic measurements alone \citep{lebreton2014a, silva-aguirre2017a}. Combining luminosities estimated from SED fitting, with oscillation frequencies, enables us to mitigate the degeneracy issue. Second, SED-fitting methods leverage multiple bandpasses of photometry whenever available, thus being robust to photometry outliers (e.g., due to stellar flares). This leads to improvements compared with the so-called direct method that combines parallaxes with single-band infrared photometry, bolometric corrections\footnote{Both SED fitting and bolometric correction tables are based on model spectra, filter transmission curves, and flux-density zero-points of individual filters.}, and reddening maps \citep[e.g.,][]{huber2017a, berger2018a, hardegree-ullman2020a}.

Asteroseismology is widely accepted as a benchmark to provide high-precision stellar ages. This method generally demands precise global seismic parameters, i.e., the frequency of maximum power ($\nu_{\rm max}$), and large frequency separation ($\Delta\nu$), which can be measured from high-precision space-based photometric time series, such as from CoRoT and \textit{Kepler}. The ongoing TESS mission is expected to provide global seismic parameters for an unprecedented number of sources. However, while it supersedes the sky coverage of its predecessors by scanning the whole sky, the majority of the targets from the first two years were observed for only 27 days. This short duration, in conjunction with lower photometric precision compared with \textit{Kepler}, leads to a lower frequency resolution, hampering $\Delta\nu$ measurements for most solar-like oscillators. Thus, with TESS data \citep{hon2021a, stello2022a}, only $\nu_{\rm max}$ can be measured at large.  Interestingly, \citet{stello2022a} found that an alternative set of stellar observables ($T_{\rm eff}$, $\nu_{\rm max}$, and radius $R$, i.e. without $\Delta\nu$) can be used to provide comparable and robust estimates of mass (random uncertainty, 12\%), and thus age (37\%). Radii derived from SED-fitting methods can improve the current precision of stellar ages, vital for understanding the formation and evolution of our Galaxy \citep[e.g.][]{silva-aguirre2018a, miglio2021a}, and for accurately dating accretion events such as the Gaia-Enceladus merger with the Milky Way \citep[e.g.][]{chaplin2020a, borre2022a}. 

Extinctions derived from SED-fitting methods are valuable, e.g., for determining the global extinctions of open clusters. Gaia astrometry and photometry has allowed the homogeneous characterisation of 2017 open clusters down to $G = 18$, including the determination of memberships \citep{cantat-gaudin2020a}. Many of these cluster members have been observed by the APOGEE \citep{donor2020a} and GALAH \citep{spina2021a} surveys. The extinction estimates of these individual spectroscopic targets allow us to precisely determine the global extinctions of the open clusters. Meanwhile, the extinction estimates of individual targets are also critical for constructing 3D dust maps for the TESS Continuous Viewing Zones (CVZs)\footnote{The CVZs center the south and north ecliptic poles and span regions with a radius of 12 degrees, i.e. 452 square degrees. The TESS light curves with a baseline of nearly 1 year are available for performing asteroseismic analysis, among others.}, as well as the \textit{Kepler} \citep{borucki2010a}, \textit{K2} \citep{howell2014a}, and PLATO \citep{rauer2014a} fields. Many of the stars in these fields are targets for searching for exoplanets. For planets in the stellar habitable zones, stellar (and therefore planetary) radius estimates are particularly important \citep{Heller2022a}. Given that asteroseismology is not generally viable for K and M dwarfs \citep{huber2019a}, Gaia parallaxes and extinctions from 3D dust maps will enable us to obtain stellar radii, which are required for characterising transiting exoplanets. 

In this work, we exploit our SED EXplorer (\texttt{SEDEX}) pipeline to estimate interstellar extinction and stellar radius from combined spectroscopic, photometric, and astrometric data, and validate the estimates with CHARA inteferometry \citep{ten-brummelaar2005a}, HST flux standards \citep{bohlin2014a}, and asteroseismology \citep{aerts2010a,basu2017a}. We focus on the stars observed by APOGEE, GALAH, and RAVE, given that their spectroscopic \teff\ values are valuable for lifting the \mbox{temperature-extinction} degeneracy. Our SED-fitting method makes use of 32 broadbands of 9 large-volume photometry databases (see Table~\ref{tab:photoParams}), exceeding previous related works  \citep[e.g.][]{berger2018a, berger2020a, queiroz2020a, steinmetz2020a}.

\begin{figure*}[!ht]
\begin{center}
\resizebox{\textwidth}{!}{\includegraphics{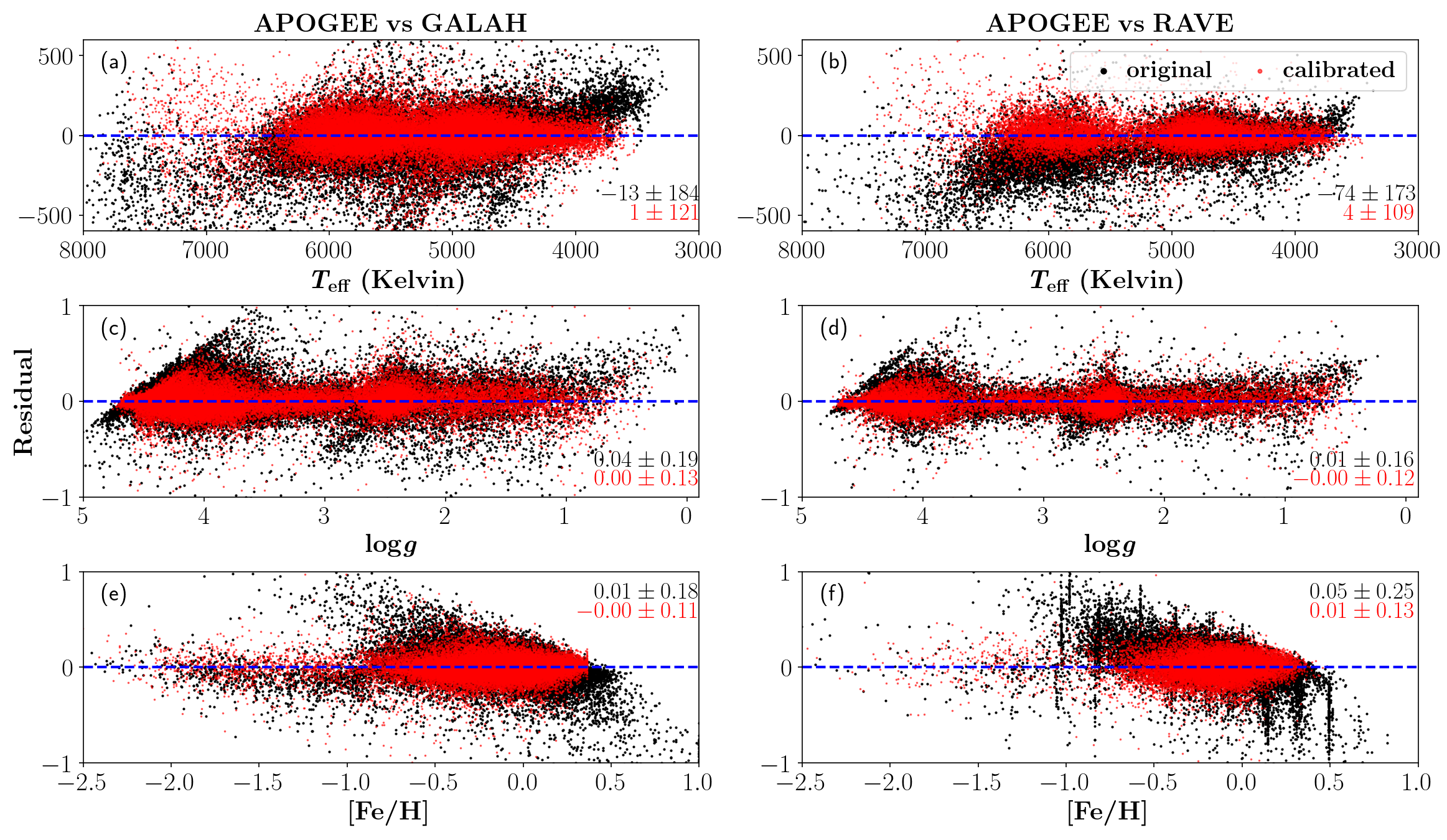}}
\caption{Comparison of the \teff\ (1st row), \logg\ (2nd row), and [Fe/H] (3rd row) values of APOGEE DR17 with those of GALAH DR3 (black, left columns) and with those of RAVE DR6 (black, right columns). The red points are similar to the black points, except that the \teff, \logg, and [Fe/H] estimates in the GALAH and RAVE catalogs are calibrated to match the APOGEE scales (see Section~\ref{homoparams}). The ordinates (residuals) are defined in the sense of APOGEE minus GALAH or APOGEE minus RAVE, whereas the abscissas correspond to GALAH or RAVE. The horizontal lines indicate perfect consistency. The mean offset and  standard deviation of the residuals are shown in each panel, with the colors having the same meaning as those of the data points.}
\label{fig:surveydatadiff}
\end{center}
\end{figure*}

\begin{figure*}
\begin{center}
\resizebox{0.9\textwidth}{!}{\includegraphics{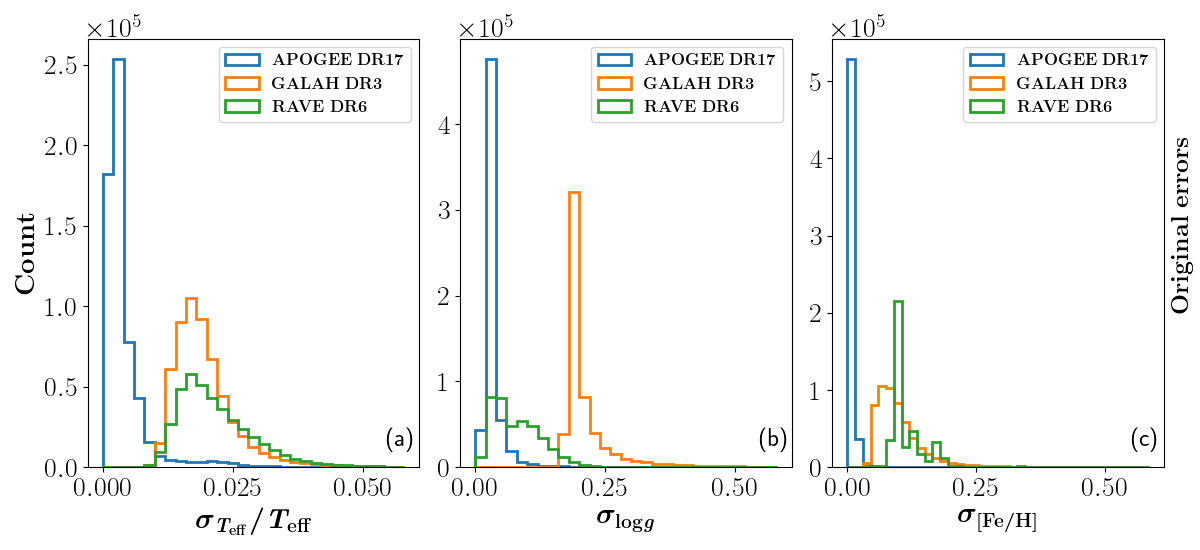}}\\
\resizebox{0.9\textwidth}{!}{\includegraphics{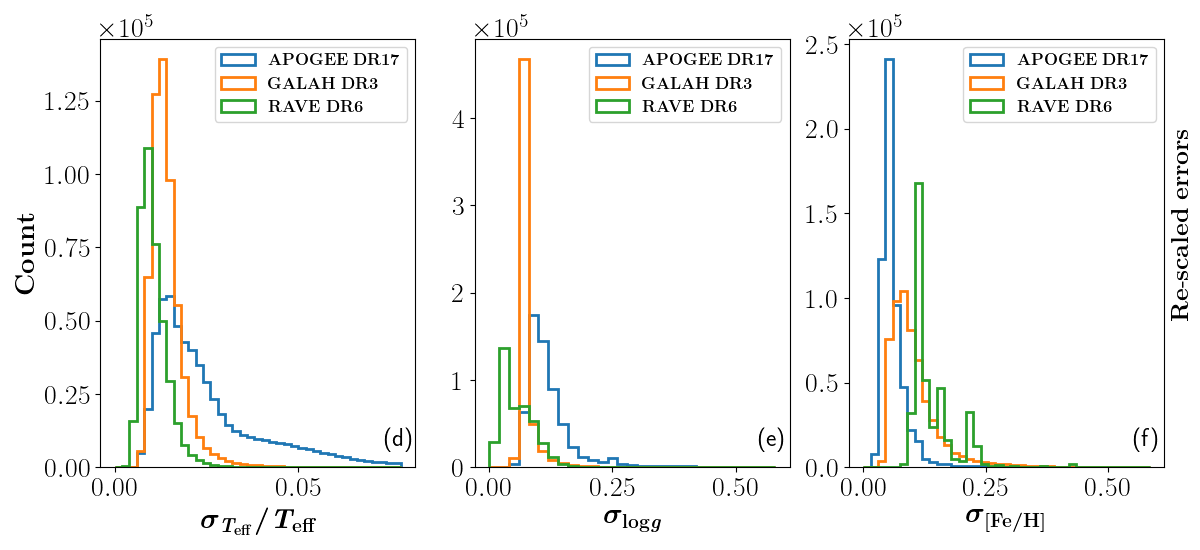}}
\caption{\textbf{Upper}: histograms of the heterogeneous uncertainties of \teff\ (left), \logg\ (middle), and [Fe/H] (right) of APOGEE DR17 (blue), GALAH DR3 (orange), and RAVE DR6 (green). \textbf{Lower}: similar to the upper panels, now  for the re-scaled uncertainties that contains the calibration uncertainties and the intrinsic uncertainties (see Section \ref{homoparams}).}
\label{fig:surveyerrors}
\end{center}
\end{figure*}

\section{Data and Methodology}

\subsection{Target Selection}\label{targetselection}
Our targets were selected from three large-volume high-resolution spectroscopic campaigns, i.e.,  APOGEE \citep[spectral resolution~$\sim22,500$,][]{wilson2019a, abdurrouf2022a}, GALAH \citep[spectral resolution~$\sim28,000$,][]{de-silva2015a,buder2021a}, and RAVE \citep[spectral resolution~$\sim7,500$,][]{steinmetz2020b, steinmetz2020a}. APOGEE has largely observed fields north of the Galactic plane, whereas GALAH and RAVE cover the southern sky. APOGEE predominantly targets red giants, aiming at tracing low Galactic latitudes owing to the near-infrared nature of the survey, while GALAH and RAVE include both dwarfs and giants, but essentially omit low Galactic latitudes according to their survey designs, to minimize contamination of unresolved multiple sources in a single fiber. APOGEE probes the distant regions of the Galaxy, with a median distance of $\sim$1.4 kpc, while GALAH and RAVE target stars that are largely nearby, with a median distance of $\sim$800 pc, as revealed by the Gaia parallax-based distances \citep{bailer-jones2021a}. Thanks to the differences in the spatial distribution and distance, APOGEE naturally probes the high-extinction space, while GALAH and RAVE explore low-extinction regions.

APOGEE DR17 contains 657,135 unique targets \citep{abdurrouf2022a}, from which we selected 616,483 stars whose \teff\ and \logg\ estimates are provided, the \texttt{STAR\_BAD} flag is unset, and Gaia EDR3 source IDs are not NaN or duplicated. GALAH DR3 contains 588,571 stars \citep{buder2021a}, from which we selected 573,593 unique stars, again, by requiring the availability of their \teff\ and \logg\ estimates, and unique Gaia EDR3 source IDs. RAVE DR6 provides stellar parameters for 451,783 unique stars \citep{steinmetz2020a}. We focused on its BD sample of 405,059 unique stars with unique Gaia DR3 source IDs. This BD sample is a subset of the RAVE DR6 database that fulfils the basic quality criterion  \texttt{algo\_conv\_madera} $\neq 1$ and uses Gaia DR2 distances for deriving stellar atmospheric parameters with the \texttt{BDASP} pipeline \citep[for more details, see Section 6 of][]{steinmetz2020a}

The entire sample selected above was further filtered by demanding that at least 5 photometric bands were available for our SED fitting (see Section~\ref{inputphotometry} for input photometry). Thus, the combined sample consists of 1,586,926 entries, among which 610,602 stars are included in APOGEE DR17, 572,702 in GALAH DR3, and 403,622 in RAVE DR6. There are 38,493 stars in common between APOGEE and GALAH, 16,499 between APOGEE and RAVE, and 30,794 between GALAH and RAVE.

\subsection{Homogenizing cross-survey atmospheric parameters}\label{homoparams}
The overlapping stars between the three surveys allow us to homogenize their atmospheric parameters. The differences between the cross-survey atmospheric parameter values are shown in Figure \ref{fig:surveydatadiff}. Substantial offsets in \teff\ are visible at low ($\sim$3800 K) and high \teff\ ($\sim$7000 K) values between APOGEE and GALAH, and at high \teff\ ($\sim$7000 K) values between APOGEE and RAVE (panels (a) and (b)). There is good agreement in \logg\ between APOGEE \& GALAH and APOGEE \& RAVE (panels (c) and (d)).  Significant systematic trends are shown in [Fe/H], where the [Fe/H] residuals between APOGEE \& GALAH and between APOGEE \& RAVE exhibit linear trends (black points, panels (e) and (f)). \bc{Moreover, vertical stripes are present in RAVE [Fe/H] (panel (f)). According to \citet{steinmetz2020a}, these stripes occur in stars with low S/N spectra, which leads to poorer fits. This can be seen in their figures 8, 9, and 10.} 

We attempted to minimize the parameter offsets by calibrating the \teff, \logg, and [Fe/H] estimates of GALAH and RAVE to match those of APOGEE. For this, we used \bc{Multilayer} Perceptrons \citep[MLPs, ][]{haykin1994a}, a supplement of feedforward neural networks where the data flows in the forward direction from input to output layer. \bc{MLP as a widely recognized algorithm of artificial neural networks is highly effective to learn non-linear relations in tabular data. The applications of MLP for regression can be found in \citet{disanto2018a}, \citet{ting2018a}, and \citet{recio-blanco2022a}, for example.} After exploring different network structures, we found the optimal architecture consists of one input layer of 10 neurons for taking features, three hidden layers with the numbers of neurons decreasing from 32, to 16 and 8, and one output layer for predicting the parameter to be calibrated. We adopted the Rectifier Linear Unit (ReLU) as the activation function and the Mean Squared Error as the loss function.

Our 10 input features are \teff, \logg, [Fe/H], Gaia EDR3 and 2MASS photometry \citep[$G$\footnote{\bc{G-band photometry used in this work was corrected following the formulae presented in \citet{riello2021a} and implemented with the \href{https://github.com/agabrown/gaiaedr3-6p-gband-correction}{code} listed in their appendix.}}, $G_{\textrm{BP}}$, $G_{\textrm{RP}}, H, K$, ][]{riello2021a, cutri2003a}, and Gaia EDR3 parallaxes \citep{lindegren2021a}. Our labels are the parameter that we intend to calibrate.  For example, for calibrating the GALAH \teff\ scale to that of APOGEE, the features include \teff, \logg, [Fe/H] in GALAH DR3, and the labels are APOGEE \teff\ values. Similarly, for calibrating the RAVE [Fe/H] scale to that of APOGEE, the features include \teff, \logg, [Fe/H] in RAVE DR6, and the labels are APOGEE [Fe/H] values. We trained 6 MLPs separately (3 MLPs for calibrating GALAH parameters and 3 for RAVE).

A comparison of our calibrated GALAH and RAVE parameters with those of APOGEE are shown in Figure~\ref{fig:surveydatadiff} (red points). The calibrated GALAH and RAVE \teff\ values are now more consistent with APOGEE, where both the bias and scatter are reduced (see the numbers in Panel (a) and (b)). The most significant improvement is in [Fe/H], where the linear trends in the residuals are removed. Furthermore, the vertical stripes are modified.

We also updated the heterogeneous uncertainties of \teff, \logg, [Fe/H] for the three surveys. Specifically, we first calculated the standard deviation $\sigma_{\teff, AG}$ of the residuals between the APOGEE \teff\ values and the calibrated GALAH \teff\ estimates ($\sigma_{\teff, AG}=$ 121 K, as given in Figure~\ref{fig:surveydatadiff}a). Similarly, we computed $\sigma_{\teff, AR}$ between APOGEE and RAVE, and  $\sigma_{\teff, GR}$ between GALAH and RAVE. To determine the typical uncertainties for APOGEE ($\sigma_{\teff, A}$), GALAH ($\sigma_{\teff, G}$), and RAVE ($\sigma_{\teff, R}$),  we assume that the three datasets are independent. Thus, we obtain 
\begin{align}
    \sigma_{\teff, AG}^2 &= \sigma_{\teff, A}^2 + \sigma_{\teff, G}^2\\
    \sigma_{\teff, AR}^2 &= \sigma_{\teff, A}^2 + \sigma_{\teff, R}^2\\ 
    \sigma_{\teff, GR}^2 &= \sigma_{\teff, G}^2 + \sigma_{\teff, R}^2.
    \label{current_rel1}
\end{align}
Solving these three linear equations yields $\sigma_{\teff, A}$, $\sigma_{\teff, G}$, and $\sigma_{\teff, R}$. Finally, we re-scaled the \teff\ uncertainties of each survey by multiplying a factor to ensure that the median value of the re-scaled \teff\ uncertainties equal to $\sigma_{\teff, A}$ for APOGEE, $\sigma_{\teff, G}$ for GALAH, and  $\sigma_{\teff, R}$ for RAVE. We updated \logg\ and [Fe/H] uncertainties for each survey using the same scheme.

The distributions of the original and re-scaled parameter uncertainties are shown in Figure~\ref{fig:surveyerrors}, and their median  uncertainties are given in Table~\ref{tab:errors}. After the scaling, RAVE stars have the most precise \teff\ and \logg\ estimates, while APOGEE stars have larger uncertainties. Note that since the re-scaled uncertainties consists of the uncertainties introduced from the calibration process, they do not necessarily represent the intrinsic uncertainties of each survey. We added a \teff\ error floor of 2.4\% in quadrature to the re-scaled \teff\ uncertainties to account for the zero point uncertainty of spectroscopic \teff\ estimates \citep{tayar2022a}\footnote{The 2.4\% \teff\ uncertainty floor is not included in the re-scaled errors reported in Table~\ref{tab:errors}.}. Thus, the systematic 2.4\% uncertainties dominate the \teff\ error budget. The homogeneous stellar atmospheric parameters and their re-scaled uncertainties (including the 2.4\% systematic \teff\ uncertainties) were subsequently used for our SED analysis.

\begin{deluxetable}{cccccccc}[t]
\tablecaption{Atmospheric parameter uncertainties of APOGEE DR17, GALAH DR3, and RAVE DR6\label{tab:errors}}
\tablewidth{0pt}
\tablehead{
\colhead{} & \multicolumn{3}{c}{re-scaled} && \multicolumn{3}{c}{original}\\
\cline{2-4} \cline{6-8}
\colhead{Survey} & \colhead{$\sigma_{\teff}$} & \colhead{$\sigma_{\logg}$} & \colhead{$\sigma_{\rm{[Fe/H]}}$} && \colhead{$\sigma_{\teff}$} & \colhead{$\sigma_{\logg}$} & \colhead{$\sigma_{\rm{[Fe/H]}}$}
}
\startdata
     APOGEE & 99 & 0.11 & 0.05 &&  12 &  0.03 & 0.008 \\
     GALAH  & 70 & 0.06 & 0.09 &&  96 &  0.19 & 0.088 \\
     RAVE   & 47 & 0.05 & 0.12 &&  97 &  0.07 & 0.095 \\
\enddata
\tablecomments{The median \teff, \logg, and [Fe/H] uncertainties of the individual catalogs before (original) and after (re-scaled) the \mbox{cross-survey} scaling (see Section~\ref{homoparams}). Note that the original cross-survey uncertainties are heterogeneous due to different definitions, while the re-scaled uncertainties consist of the calibration and intrinsic uncertainties. }
\end{deluxetable}

\begin{deluxetable*}{cccccccccc}
\tablecaption{Photometric system parameters\label{tab:photoParams}}
\tablewidth{0pt}
\tablehead{
\colhead{Photometric System} & \colhead{Filter} & Mag system& \colhead{FTC} & \colhead{$\lambda_{\rm P}$} & \colhead{$\overline{f}_{0, \lambda}$} & \colhead{$\overline{f}_{0, \nu}$} & \colhead{Ref.} & \colhead{$m_{0}$} & \colhead{Ref.}\\
\colhead{} & \colhead{} & \colhead{} &\colhead{}  & \colhead{\AA} & \colhead{erg/s/cm$^2/$\AA} & \colhead{Jy} & \colhead{} & \colhead{mag} & \colhead{}}
\decimalcolnumbers
\startdata
Hipparcos  & $H_P$  & Vega & $\lambda T$  &  5586  &    3.296E-9   & -- &  1 & 0 & 1 \\
\hline
Tycho2     & $B_T$  & Vega & $\lambda T$  &  4220  &    6.798E-9   & -- & 1 & 0 & 1 \\
          & $V_T$   & Vega & $\lambda T$  &  5350  &    4.029E-9   & -- & 1 & 0 & 1 \\
\hline            
          & $G_{BP}$ & Vega   & $T$            &  5109.7  &    4.110E-9 & -- & 2 & 0 & 2 \\
Gaia EDR3  & $G$     & Vega   & $T$            &  6217.6  &    2.536E-9 & -- & 2 & 0 & 2 \\
          & $G_{RP}$ & Vega   & $T$            &  7769.1  &    1.299E-9 & -- & 2 & 0 & 2 \\          
\hline 
          & $B$    & Vega   & $\lambda T$    &  4368.4  &    6.459E-9 &   -- & 1 & 0 & 1 \\        
          & $V$    & Vega   & $\lambda T$    &  5486.2  &    3.735E-9 &   -- & 1 & 0 & 1 \\        
APASS      & $g$   &AB   & $T$               &  4702.5  &   --        & 3631 & 3 & 0.003 & 8 \\
          & $r$    &AB   & $T$               &  6175.6  &   --        & 3631 & 3 & $-0.006$ & 8 \\
          & $i$    &AB   & $T$               &  7490.0  &   --        & 3631 & 3 & $-0.016$ & 8  \\
\hline           
          & $u$    &AB & $T$          &  3556.5  &    --   & 3631 & 3 & 0.037  & 8 \\
          & $g$    &AB & $T$          &  4702.5  &    --   & 3631 & 3 & $-0.010$ & 8 \\
SDSS       & $r$   &AB & $T$          &  6175.6  &    --   & 3631 & 3 & 0.003 & 8 \\
          & $i$    &AB & $T$          &  7490.0  &    --   & 3631 & 3 & $-0.006$ & 8 \\
          & $z$    &AB & $T$          &  8946.7  &    --   & 3631 & 3 & $-0.016$ & 8 \\
\hline            
          & $g$      &AB & $\lambda T$    &  4814.1  &    -- & 3631 & 4 & 0 & 4 \\
          & $r$      &AB & $\lambda T$    &  6174.3  &    -- & 3631 & 4 & 0 & 4 \\
Pan-STARRS & $i$     &AB & $\lambda T$    &  7515.8  &    -- & 3631 & 4 & 0 & 4 \\
          & $z$      &AB & $\lambda T$    &  8663.6  &    -- & 3631 & 4 & 0 & 4 \\
          & $y$      &AB & $\lambda T$    &  9616.9  &    -- & 3631 & 4 & 0 & 4 \\
\hline    
          & $u$      &AB & $T$    &  3493.36 &  -- & 3631 & 5 & 0 & 5 \\
          & $v$      &AB & $T$    &  3835.93 &  -- & 3631 & 5 & 0 & 5 \\
SkyMapper & $g$      &AB & $T$    &  5075.19 &  -- & 3631 & 5 & 0 & 5 \\
          & $r$      &AB & $T$    &  6138.44 &  -- & 3631 & 5 & 0 & 5 \\
          & $i$      &AB & $T$    &  7767.98 &  -- & 3631 & 5 & 0 & 5 \\
          & $z$      &AB & $T$    &  9145.99 &  -- & 3631 & 5 & 0 & 5 \\
\hline  
          & $J$      & Vega   & $\lambda T$    & 12350  & 3.129E-10 & -- & 6 & $-0.018$ & 8 \\
2MASS      & $H$     & Vega   & $\lambda T$    & 16620  & 1.133E-10 & -- & 6 & 0.035 & 8 \\
          & $K_{S}$  & Vega   & $\lambda T$    & 21590  & 4.283E-11 & -- & 6 & $-0.014$ & 8 \\ 
\hline            
          & $W1$      & Vega   & $\lambda T$   &  33526  &  8.179E-12 & --  & 7 & 0 & 7 \\
          & $W2$      & Vega   & $\lambda T$   &  46028  &  2.415E-12 & --  & 7 & 0 & 7 \\
ALLWISE   & $W3$      & Vega   & $\lambda T$   & 115608  &  6.515E-14 & --  & 7 & 0 & 7 \\
          & $W4$      & Vega   & $\lambda T$   & 220883  &  5.090E-15 & --  & 7 & 0 & 7 \\
\enddata
\tablecomments{Column 4 indicates whether each filter transmission curve (FTC) from the original source has been multiplied by $\lambda$. Pivot wavelengths, $\lambda_{\rm p}$, given in column 5 are calculated from the adopted FTCs when they are unavailable from literature. Since the zero-point flux density values ($\overline{f}_{0, \lambda}$ ) for the Gaia system are not available from literature, we calculated it using the Vega spectrum, \texttt{alpha\_lyr\_mod 002.fits}. This spectrum is adopted from the CALSPEC Calibration Database, and rescaled to set the flux equal to $f_{550}$ = 3.62286 10$^{-11}$ W m$^{-2}$ nm$^{-1}$ at the wavelength $\lambda$ = 550.0 nm, which is assumed as the flux of an unreddened A0V star with $V$ = 0   \citep[also see][]{riello2021a}. The references of the FTCs and zero points ($\overline{f}_{0, \nu}$ and $m_{0}$) are given in columns 8 and 10, respectively. 1: \citet{mann2015a}; 2: \citet{riello2021a}; 3: SDSS webset at \url{http://classic.sdss.org/dr7/instruments/imager/index.html\#filters}; 4. \citet{tonry2012a}; 5: \citet{bessell2011a}; 6: \citet{cohen2003a}; 7: \citet{jarrett2011a}; 8: \citet{casagrande2014a}.}
\end{deluxetable*}

\subsection{Input Photometry}\label{inputphotometry}
Observed SEDs were constructed from 32 bandpasses of 9 photometric databases, with the following photometric systems when available: Gaia DR3 \citep{riello2021a}, PanSTARRS DR1\citep{chambers2016a}, APASS DR9 \citep{henden2016a}, Tycho2 \citep{hog2000a}, Hipparcos \citep{van-leeuwen2007a}, SkyMapper DR2 \citep{onken2019a}, SDSS DR13 \citep{alam2015a}, 2MASS \citep{cutri2003a}, and ALLWISE \citep{cutri2013a}. These data sets collectively span a broad range in wavelength, from the optical through the infrared. Generally, the surveys provide measurements averaged over multi-epoch photometry, such as Gaia photometry. The use of average measurements helps to reduce the scatter due to stellar variability. We discarded ALLWISE photometry in the W3 and W4 bands, because of potential biases of their zero-point calibration \citep[see][]{yu2021a}, large photometric uncertainties, and possible contamination by warm interstellar dust \citep{davenport2014a}. We combined the various photometric and spectroscopic datasets, based on their Gaia EDR3 source IDs. For the photometry, these are provided by Gaia \citep{marrese2019a}, while the spectroscopic surveys provide their internal crossmatch to Gaia.

We added uncertainty floors in quadrature to the formal magnitude uncertainties of individual photometric catalogs: 0.02 mag for GAIA, Hipparcos, Tycho2, and 2MASS, 0.03 mag for ALLWISE, and 0.06 mag for SDSS, PanSTARRS, APASS, and SkyMapper \citep[also see][]{eastman2019a,godoy-rivera2021a}. We consider this step necessary. First, the SED fitting would otherwise be exclusively controlled by Gaia photometry and less sensitive to the other photometric measurements because of the very small formal uncertainties in Gaia photometry. Second, higher weights should be assigned to Gaia, Hipparcos, Tycho2, 2MASS, and ALLWISE photometry, due to their superior photometric precision for our sample. Lastly, adding uncertainty floors mitigates potential problems associated to the not well understood photometric zero-points. These are known to introduce additional uncertainties of up to 1-2\% for ground-based photometry, compared with space-based spectrophotometry from HST/STIS \citep{bohlin2014a}. In case formal uncertainties are unavailable, the uncertainty floors were adopted but inflated by a factor of three.

To account for potential photometric outliers (e.g. due to flares or photometric saturation), we fitted a black-body distribution to each observed SED, assuming zero extinction. This step was performed iteratively to detect and remove one outlier photometric measurement at each iteration, until none of the flux densities deviated by more than 30\% from the best-fitting black-body spectrum. This homogeneous method applied to heterogeneous photometry is preferable over filtering data with complex combination of photometric quality flags.  During this step, magnitudes were converted to flux densities using the absolute calibration of the flux densities of the individual filters listed in Table~\ref{tab:photoParams}. Next, we computed model flux densities calculated from stellar spectral libraries.

\subsection{Model Spectra}
Several stellar spectral libraries have been developed for analysing observed spectra, such as ATLAS \citep{kurucz1979a}, MARCS \citep{gustafsson2008a}, PHOENIX \citep{husser2013a}, BOSZ \citep{bohlin2017a}, among others. While all these models have been widely used in literature for SED fitting, we chose MARCS in this work for consistency, because this library was adopted for spectrum analyses to determine stellar parameters and chemical abundances by APOGEE \citep{jonsson2020a}  GALAH \citep{buder2021a}, and RAVE \citep{steinmetz2020a}. We refer the reader to \citet{gustafsson2008a} and the website \hyperlink{https://marcs.astro.uu.se/}{https://marcs.astro.uu.se/} for details on MARCS. We note that the maximum wavelength of the spectra in MARCS is 20~$\mu$m, which  leads to missing flux when calculating bolometric flux for late type stars. For this reason, we extrapolated MARCS spectra in logarithmic scale with a cubic polynomial, out to 30 $\mu$m \citep{yu2021a}.

Given that APOGEE DR17 contains numerous B and A type stars, and the maximum \teff\ is 8000 K in the MARCS grid, we adopted the BOSZ library for modelling hotter main-sequence (MS) stars (\teff~$>$~8000 K), complementing the MARCS library for modelling cooler stars (\teff~$\leq$~8000 K). We note that there is a negligible systematic difference (sub 1\%) in the derived radii with the two libraries at this \teff\ boundary.

To compute the model flux density, we convolve each filter transmission curve (see the references given in Table \ref{tab:photoParams}) with MARCS model spectra. All the transmission curves, which are in coarse wavelength intervals, have been interpolated to the higher model wavelength resolution ($\lambda/\Delta\lambda=20,000$) in order not to skip any line features \citep{bessell2012a}.

\subsection{General Fitting}\label{fittingmethod}
The first step of the fitting process was to search for the best-fitting spectral model without spectral model interpolation. For each star, we began by seeking the MARCS models whose metallicities were closest to the observed one, adopted from the same reference as \teff\ and \logg\ (Sect.~\ref{targetselection}). When observed metallicity estimates are unavailable, solar metallicity was assumed. From these models we chose the models with the 4 closest grid \teff\ values with respect to the observed \teff, and further picked the models with the 4 closest grid \logg\ values compared with the observed \logg. This led to a maximum of 16 models, if available, in the \teff-\logg\ plane, bracketing the observed values. The best-fitting model was the one with the minimum reduced $\chi ^2$ in flux density. To avoid numerical overflow issues in the minimisation step, the flux densities and their uncertainties were converted to a logarithmic scale. 

The second step was to use the best-fitting model to further remove photometric outliers of the observed SED prepared in Sect.~\ref{inputphotometry} in an iterative way. This procedure is similar to the initial cleaning step introduced in Section Sect.~\ref{inputphotometry} to refine the input photometry, except here, we fitted SEDs rather than a black-body. To automatically remove photometric outliers of an observed SED, we calculated the relative difference between the observed and the best-fitting model flux densities, and rejected the measurements if they were greater than a threshold. This procedure was repeated until either no more outliers were found or the measurements were too sparse to guarantee a reliable fitting. For the relative difference, we adopted a threshold of 10\%, and a number of valid photometric measurements no less than 5. Only the fits passing these criteria were retained for the subsequent analysis.

The third step was to estimate stellar parameters and their uncertainties by fitting the pruned SEDs with interpolated models in a Bayesian approach. First, we linearly interpolated the pre-selected 16 models to obtain the flux densities in logarithmic scale in the 32 bandpasses, with a grid resolution of 5 K in \teff\, and 0.25 dex in \logg. 
Then, we used normal priors for \teff\ and \logg, which are centred at observed values with the standard deviations equal to observed uncertainties. Our fitting model is 
\begin{equation}
F_{\lambda, \rm obs} = (R/d)^2 10^{-0.4\beta(\lambda)A_{V}} F_{\lambda, \rm mod} (\teff, \logg, \rm{[Fe/H]}),
\end{equation}
where $F_{\lambda, \rm obs}$ and $F_{\lambda, \rm mod}$ are the observed and model flux densities, respectively, $R$ is stellar radius,  $d$ is heliocentric distance, $\beta(\lambda)$ is an extinction law as a function of wavelength, and $A_V$ is extinction in $V$. Our likelihood function was assumed to be Gaussian, and was optimized with the Levenberg-Marquardt optimization algorithm.  Finally, we fitted a Gaussian to each posterior and used its mean and standard deviation to estimate a parameter (e.g., extinction) and its uncertainty, respectively. We refer the reader to \citet{yu2021a} for examples of SED fits.

We note that extinction is a direct output of our pipeline. Bolometric flux is calculated by integrating the best-fitting spectrum, and luminosity is derived by combining bolometric flux and distance. Radius is computed from luminosity and input \teff, and angular radius is inferred from radius and distance. We adopted photogeometric distances from \citet{bailer-jones2021a} whenever available, and their geometric distances otherwise. The uncertainties in  extinction and bolometric flux are estimated from the standard derivations of the Gaussians fitted to the posteriors. The uncertainties in luminosity, radius, and angular radius are obtained via error propagation.

Our SED fitting barely depends on the choice of the following general extinction laws as long as using $R(V)=3.1$ (adopted in our work): CCM89 \citep{cardelli1989a}, O94 \citep{odonnell1994a}, F99 \citep{fitzpatrick1999a}, F04 \citep{fitzpatrick2004a}, M14 \citep{maiz-apellaniz2014a}, G16 \citep[][reducing to the F99 model with $f_A$=1.0]{gordon2016a}, F19 \citep{fitzpatrick2019a}. We tested the difference between these models by running the fitting for a sample of $\sim$7000 asteroseismic targets including dwarfs and giants \citep{serenelli2017a, pinsonneault2018a} and found that the systematic differences in radius and extinction are within the formal uncertainties. This is because the extinction model differences are only significant in the near infrared. For example, $A_{\lambda}/A_V$ can be different by $\sim$10\% for 2MASS $K_S$ band. However, except for very cool stars (\teff~$<$~3300 K), the infrared flux does not significantly contribute to the SED, making the extinction laws essentially indistinguishable. For this reason, we adopted the F19 model for our final analysis. 

Our fitting allows for a negative extinction estimate for each interpolated model. This accounts for the fact that extinction is coupled with effective temperature, which can be biased by several 100~K. We removed stars with $A_V$ uncertainties not compatible with being positive at 2$\sigma$. Allowing for negative extinction thus enables us to analyze low-extinction stars. Our final catalog consists of 1,566,810 entries, among which 1,484,987 are unique stars. The numbers of these entries from APOGEE DR17, GALAH DR3, and RAVE DR6 are 593,374, 571,358, and 402,078, respectively. The homogeneously calibrated atmospheric parameter estimates, their re-scaled uncertainties, and the derived parameter values are given in Table~\ref{tab:catalog}.

\section{Validation}\label{validation}
\begin{figure}
\begin{center}
\resizebox{\columnwidth}{!}{\includegraphics{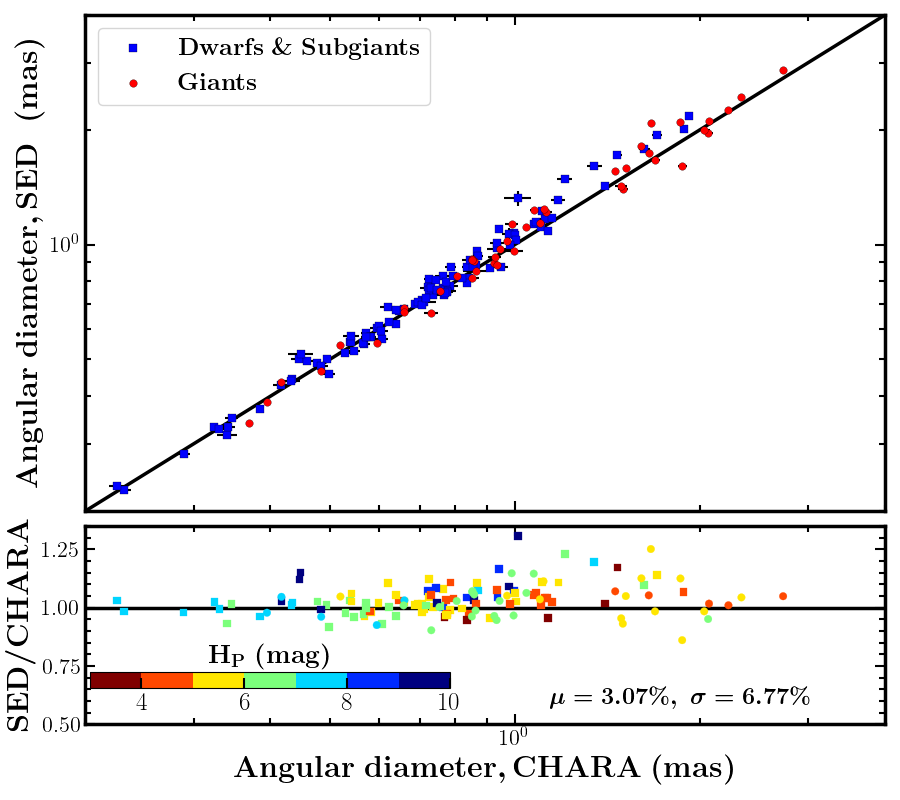}}
\caption{Comparison of angular diameters from the SED fitting to those from CHARA interferometry. \textbf{Top}: Angular diameters compared for the 83 dwarfs \& subgiants (\logg~$>$~3.5, blue squares) and the 33 giants (\logg~$\leq$~3.5, red dots). The one-to-one line (perfect match) is shown in black. \textbf{Bottom}: Angular diameter ratio (SED/CHARA) shown for the same set of populations, but color coded by Hipparcos $H_P$ magnitude. The mean and the standard deviation of the ratios are also labelled.}
\label{fig:angulardiametervalid}
\end{center}
\end{figure}

\begin{figure}
\begin{center}
 \resizebox{\columnwidth}{!}{\includegraphics{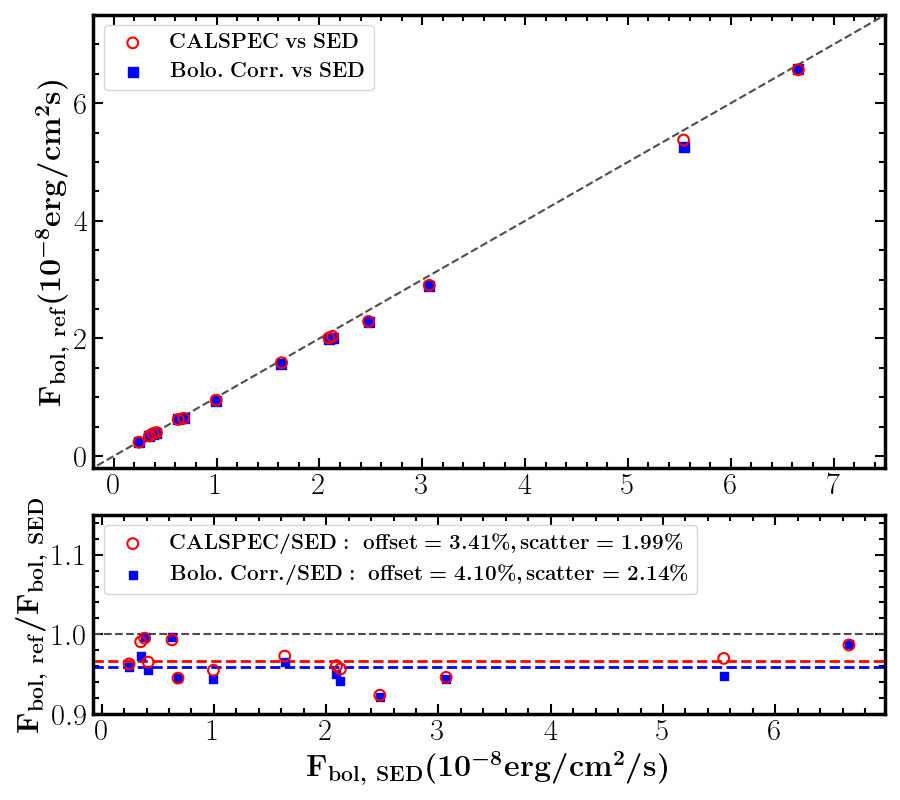}}
\caption{\textbf{Upper panel}: Comparison of bolometric flux derived from the SED fitting with that computed from the HST CALSPEC/STIS spectrophotometry of flux standards \citep[red circles,][]{bohlin2014a}, and with that calculated from bolometric corrections \citep[blue squares,][]{casagrande2018a}. The dashed line marks the one-to-one relation. \textbf{Lower panel}: Bolometric flux ratio as a function of our bolometric flux estimates. The red and blue dashed lines show the mean ratios, while the grey dashed line denotes perfect one-to-one agreement. The numbers in the legend indicate the offsets and the scatters.}
\label{fig:bolometricfluxvalid}
\end{center}
\end{figure}

\subsection{CHARA Interferometry}\label{chara}
We compiled a sample of 180 dwarfs and giants that have been observed by the CHARA interferometer \citep{baines2010a, boyajian2012a, boyajian2013a, maestro2013a, von-braun2014a, kane2015a, boyajian2015a, ligi2016a, white2018a, karovicova2020a, karovicova2022a, karovicova2022b}.  This sample was further pruned for validation, by requiring that there are at least 4 optical photometric measurements ($\lambda<1~\mu$m) retained after outlier-photometry clipping for the SED fitting. Since CHARA stars are preferentially bright, this criterion essentially requires the availability of Gaia $G$, Hipparcos $H_P$, and Tycho2 $B_T$ and $V_T$ photometry. We then performed the SED fitting using interferometric \teff\ and spectroscopic \logg\ values as priors (taken from the aforementioned references).

The angular diameter comparison shown in Figure~\ref{fig:angulardiametervalid} yields good agreement, with an offset of 3.07\% (SED/CHARA), and a scatter of 6.77\%. Inspecting the stars with angular diameters in the range \mbox{1~$<\theta/\rm{mas}<$~3}, particularly for dwarfs/subgiants (blue squares in the top panel), suggests that our angular diameters could be overestimated. This overestimation is probably caused by saturated photometry, since the dwarfs/subgiants in this angular diameter range are very bright. We note that this offset is smaller than that found by \citet{tayar2022a}, who compared interferometric angular diameters from CHARA with different beam combiners, yielding a systematic median difference of 4\%. 

\subsection{HST spectrophotometry}\label{hst}
The best way to validate bolometric flux so far probably have been using HST/STIS CALSPEC spectrophotometry. There is evidence to suggest that its \texttt{relative} fluxes from the visible to the near-IR wavelength of ~2.5 $\mu$m are currently precise to $\sim$1\% for the primary reference standards \citep{bohlin2014a}. Meanwhile, the bolometric corrections serve as an alternative approach to estimate the bolometric flux \citep{casagrande2018a}. We use the same sample of the HST/STIS CALSPEC primary flux standards as \citet{casagrande2018a} for our validation.

Figure~\ref{fig:bolometricfluxvalid} compares our bolometric flux with that calculated from the CALSPEC spectrophotometry, revealing a scatter of 1.99\%. This small scatter demonstrates that our SED fitting is precise, given that the CALSPEC spectrophotometry is precise to $\sim$1\%. We notice a somewhat large offset of 3.41\%, in the sense that our bolometric flux is larger. The reason for this offset is uncertain, given that the CALSPEC spectrophotometry was largely well calibrated to reach high precision. We note that, though \citet{casagrande2018a} attained a sub-1\% offset between their bolometric flux scale derived from bolometric corrections and those from CALSPEC spectrophotometry, their scale was based on their adopted value of solar absolute magnitude. As pointed out by \citet{casagrande2018a}, solar absolute magnitude is an arbitrary zero-point and any value is equally legitimate on the condition that once chosen, all bolometric corrections are scaled accordingly. On the other hand, an independent comparison of luminosities between the SED fitting and asteroseismology reveals a sub-1\% offset (see next Sect.). This suggests that our bolometric flux scale is not necessarily the root cause to the 3.41\% offset, and any bias of the absolute scale of CALSPEC spectrophotometry, if exists, could contribute to this offset.

\begin{figure*}
\begin{center}
\resizebox{1.05\columnwidth}{!}{\includegraphics{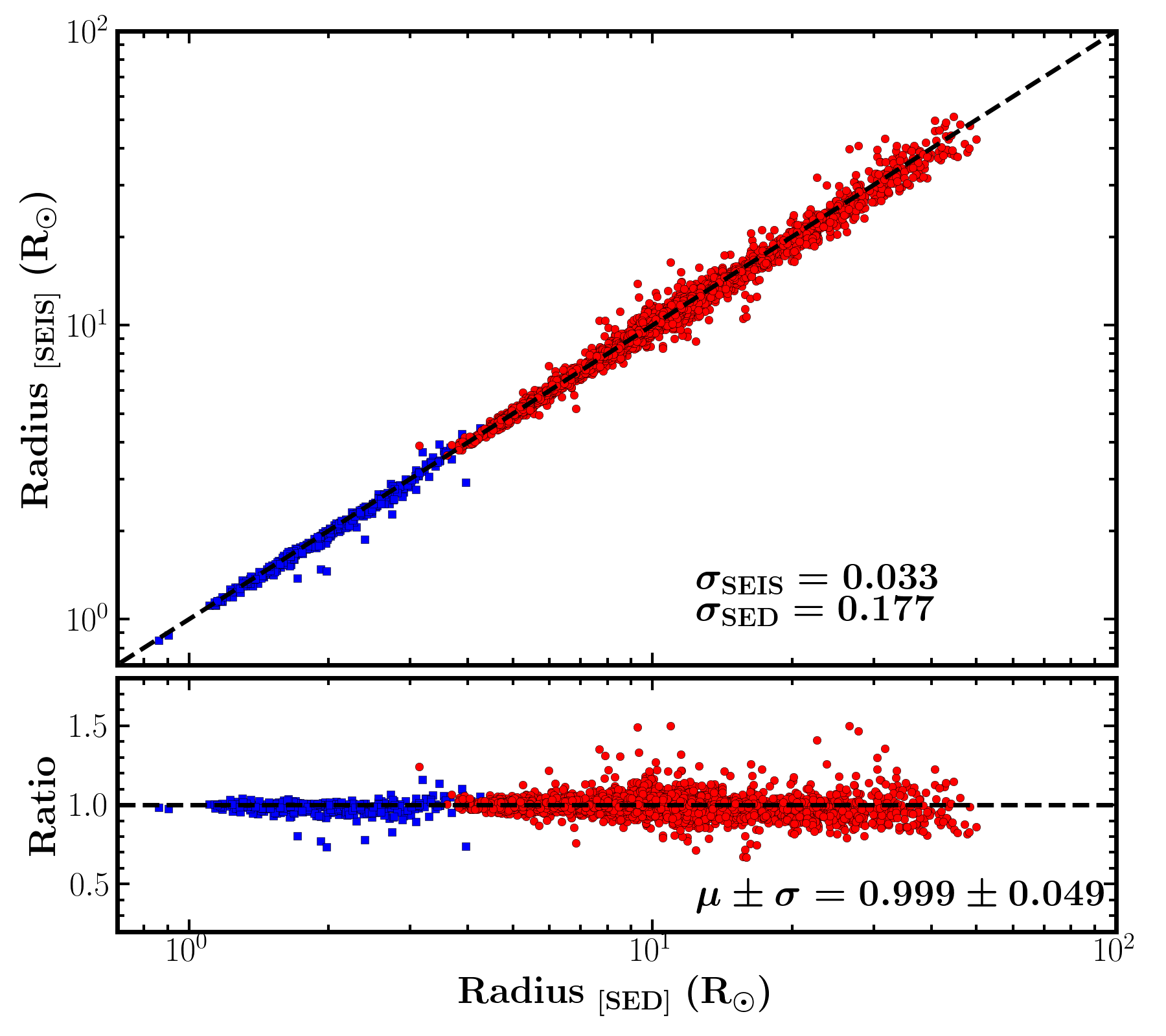}}
\resizebox{1.05\columnwidth}{!}{\includegraphics{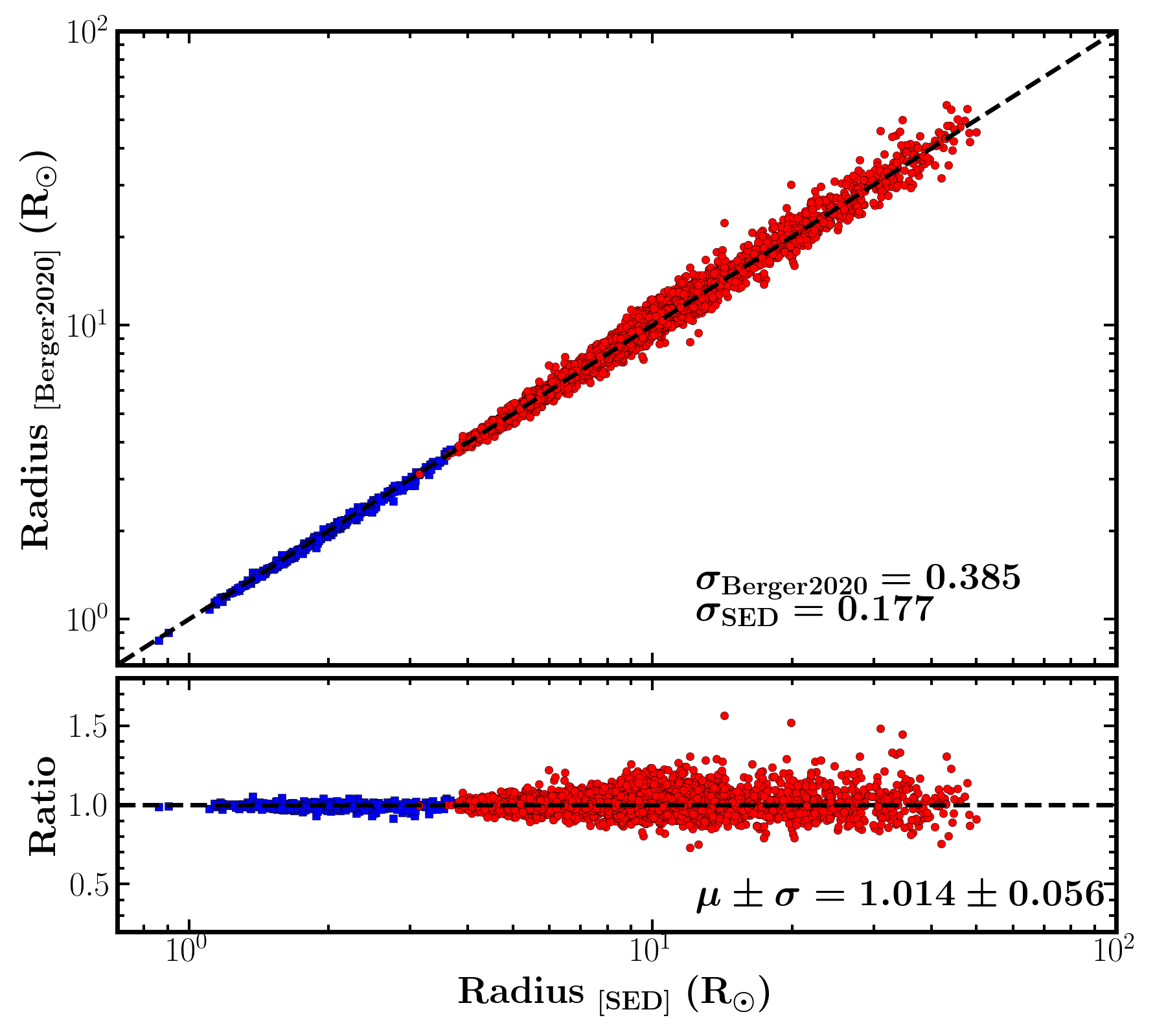}}
\caption{\textbf{Left}: Comparison of radii determined from the SED fitting with those from asteroseismology. The blue squares indicate dwarfs and subgiants from \citet{serenelli2017a}, while the red dots indicate giants from \citet{pinsonneault2018a}. The median formal uncertainties of the two sets of radii are shown in the top panel, and the mean ratio of the seismic to SED-fitting radii and its standard deviation are given in the bottom panel. The dashed lines represent perfect consistency. \textbf{Right}: Similar to the left panels except for the seismic radii replaced by those from \citet{berger2020a}.}
\label{fig:radiivalid}
\end{center}
\end{figure*}

\subsection{Asteroseismology}\label{validseis}
In this section, we validate our estimates of radius, luminosity, and extinction with asteroseismic counterparts from the APOKASC catalogs for dwarfs \& subgiants \citep{serenelli2017a}, and for giants \citep{pinsonneault2018a}. These two studies leveraged asteroseismic constraints with SDSS \textit{griz} photometry and APOGEE spectroscopy to determine stellar parameters, using a grid-based modelling method for dwarfs and subgiants \citep{serenelli2017a}, and an empirical method for giants \citep{pinsonneault2018a}. Their robust zero-point calibration of stellar parameters enables us to test the accuracy and precision of our results. We also compare our parameter estimates with the latest, homogeneous \textit{Kepler} stellar properties catalog \citep{berger2020a}, which was based on a grid-based modelling approach, independent of asteroseismology. It is important to note that for our SED fitting we used the same \teff, \logg, and metallicity values as \citet{serenelli2017a}, \citet{pinsonneault2018a}, and \citet{berger2020a} did, rather than the latest catalog we compiled in Section 2.1, to eliminate the systematics caused by different atmospheric parameter scales.

\subsubsection{Radii}
Figure~\ref{fig:radiivalid} shows good consistency between radii derived from our SED fitting and those from asteroseismology, with an offset of 0.1\% and a scatter of 4.9\%. The comparison between our radius estimates and those from the \textit{Kepler} stellar property catalog \citep{berger2020a} also yields good agreement, with an offset of 1.4\% and a scatter of 5.6\%. In both comparisons, we find a smaller dispersion in dwarfs (blue dots) than in giants (red dots).  This is in line with the fact that the giants are more distant compared to the dwarfs, and are thus subject to larger uncertainties in parallax and distance.
 
Our independent radius estimates provide an opportunity to test how well the seismic scaling relations have been empirically calibrated  by \citet{pinsonneault2018a}. We note that the asteroseismic radius scale of red giants by \citet{pinsonneault2018a} was calibrated to match the fundamental measurement of the mean mass of red-giant-branch (RGB) eclipsing binary stars in two open clusters. As \citet{pinsonneault2018a} pointed out, their asteroseismic radius scale for core-Helium-burning (CHeB) stars should be used with caution, because the uncertain mass loss on the RGB complicates an absolute radius calibration for CHeB stars. Thus, their CHeB radius scale was assumed to be the same as that for RGB stars.  Since our radius scale is not affected by population type (RGB or CHeB), the strong one-to-one correlation between the radii from \citet{pinsonneault2018a} and our measurements, would suggest high degree of self-consistency in their asteroseismic radius scale.

\begin{figure*}
\begin{center}
\resizebox{1.05\columnwidth}{!}{\includegraphics{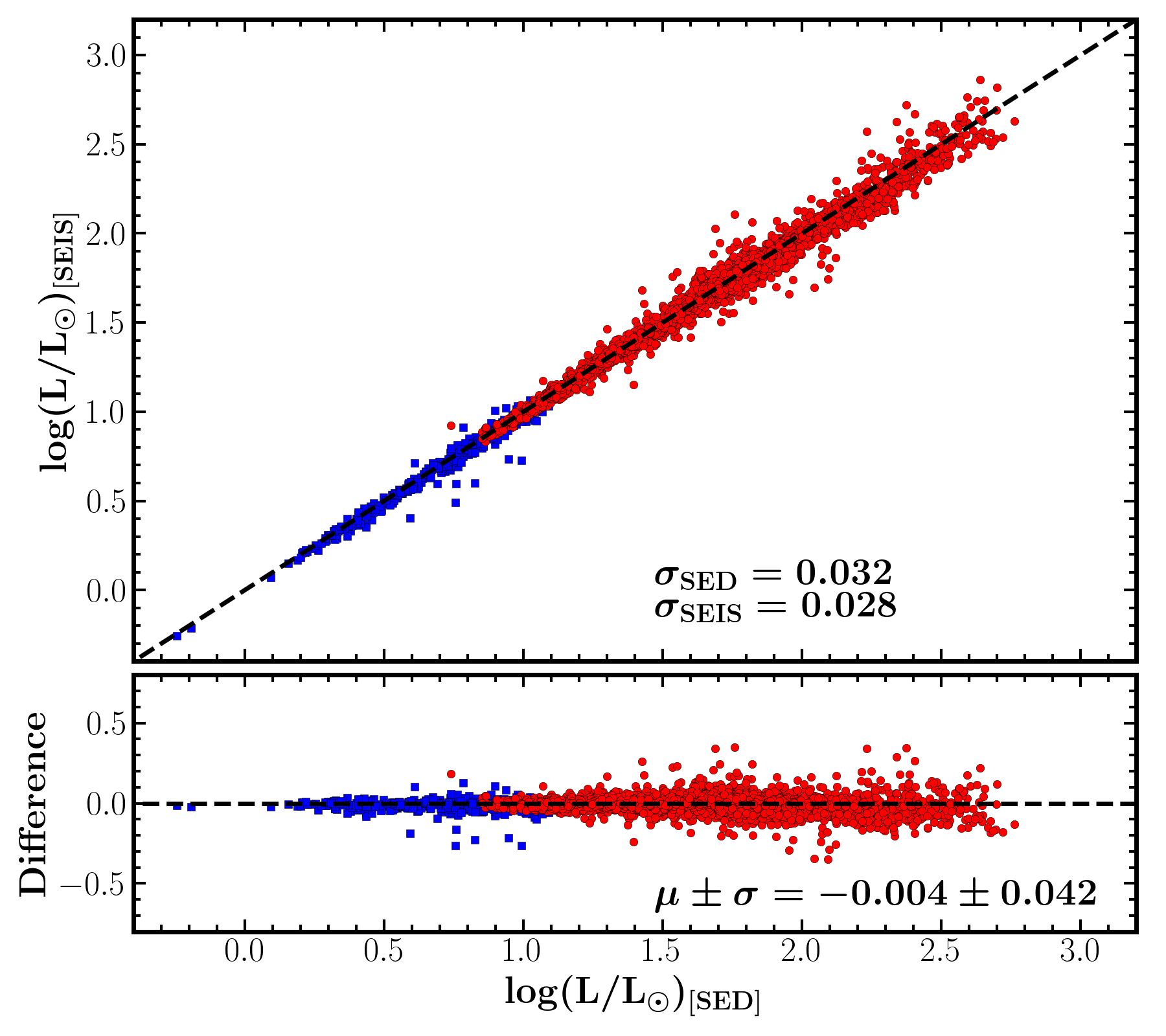}}
\resizebox{1.05\columnwidth}{!}{\includegraphics{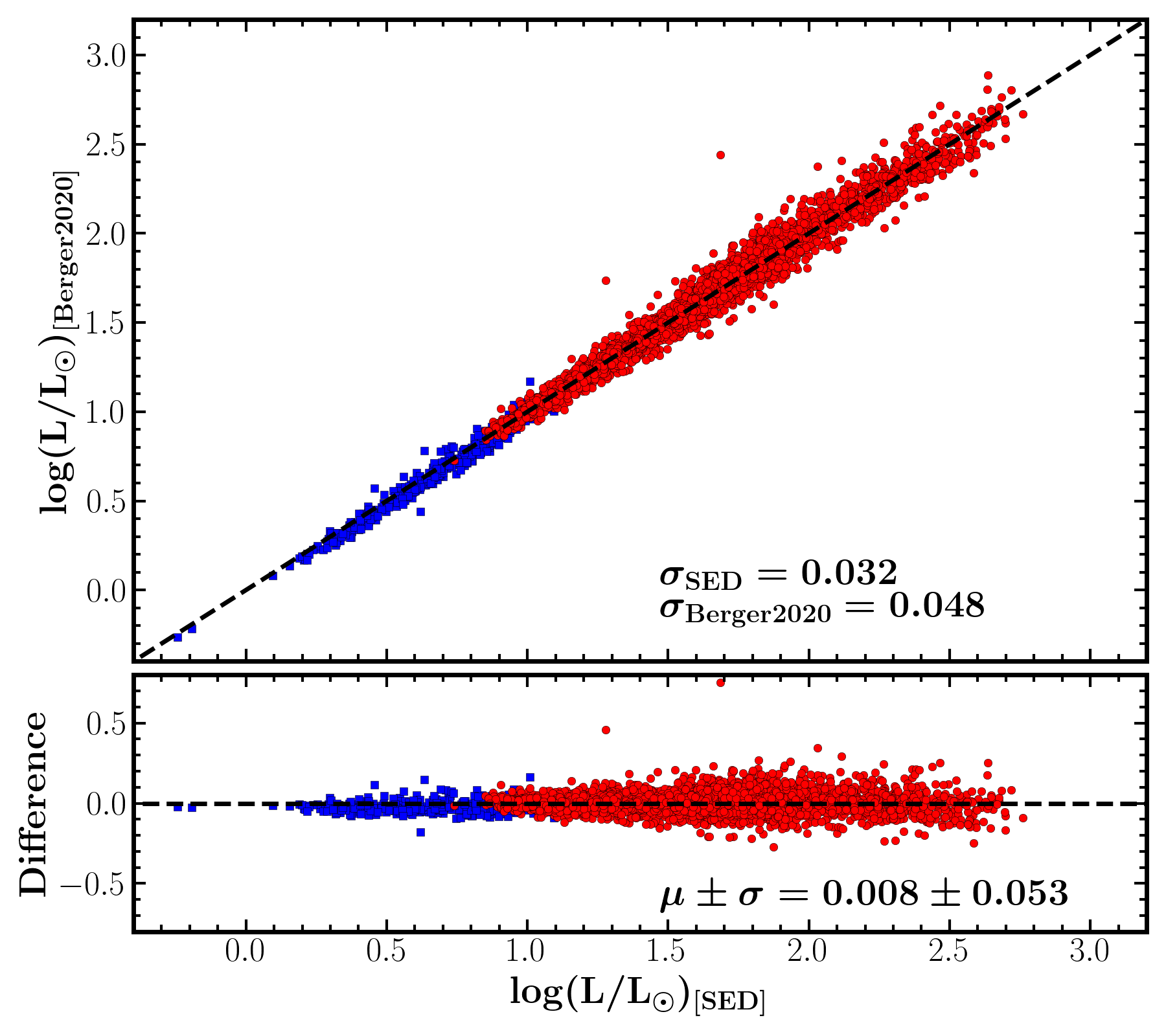}}
\caption{\textbf{Left}: Comparison of luminosity determined from the SED fitting with those from asteroseismology. The symbols, line, and text have the same meaning as in Figure~\ref{fig:radiivalid}, now for luminosity. \textbf{Right}: Similar to the left panels, except for the seismic luminosity replaced by those from \citet{berger2020a}. Our SED luminosities are computed from our bolometric fluxes and Gaia based distances \citep{bailer-jones2021a}, while the luminosities from asteroseismology and \citet{berger2020a} were derived from their radii and temperatures.}
\label{fig:lumivalid}
\end{center}
\end{figure*}

\begin{figure}
\begin{center}
\resizebox{1.0\columnwidth}{!}{\includegraphics{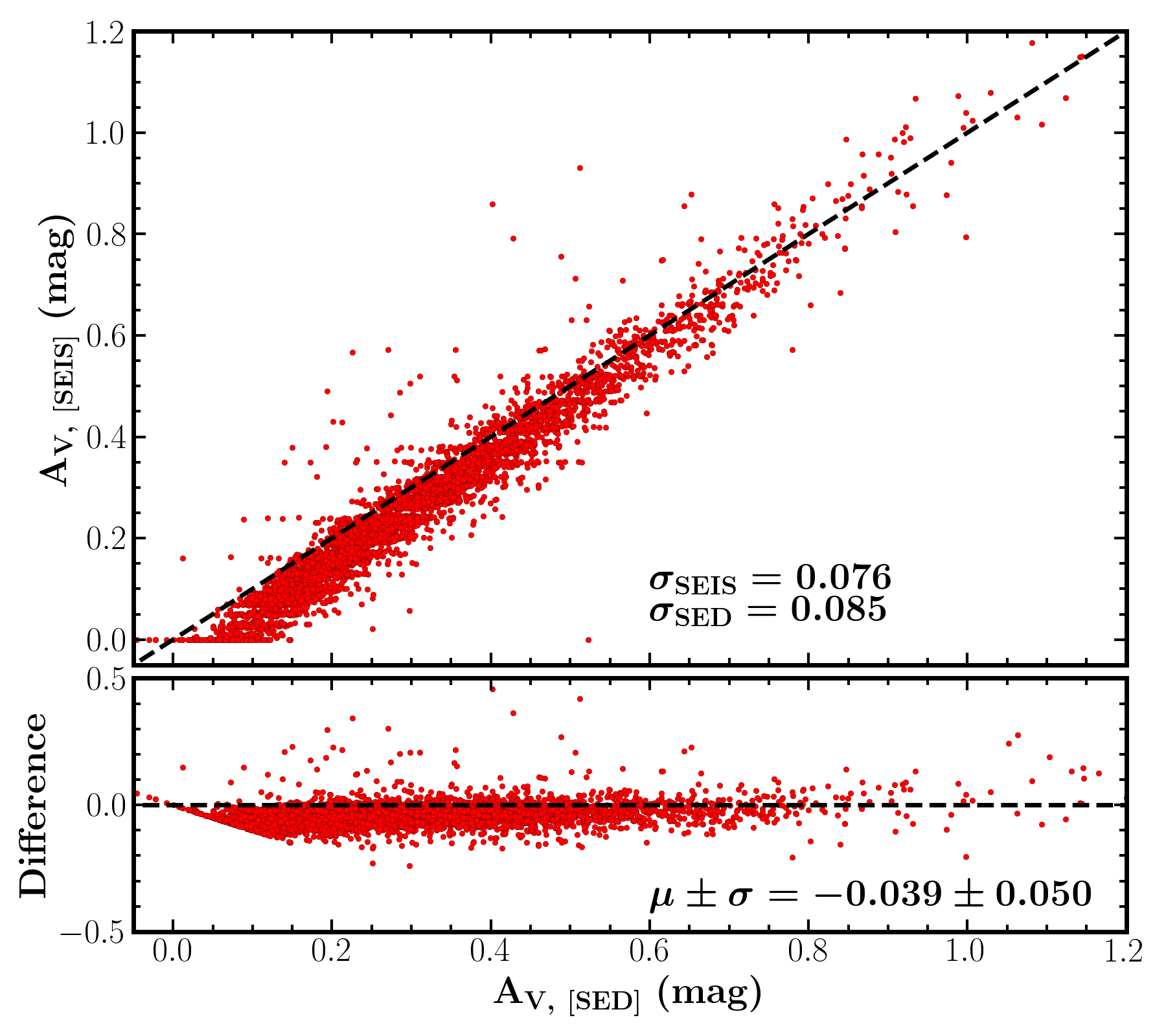}}
\caption{Comparison of extinctions determined from the SED fitting with those from asteroseismology. The symbols, line, and text have the same meaning as in Figure~\ref{fig:radiivalid}.  No extinctions were provided for dwarfs and subgiants by \citet{serenelli2017a}.}
\label{fig:Avvalid}
\end{center}
\end{figure}

\begin{figure*}[t]
\begin{center}
\resizebox{0.9\textwidth}{!}{\includegraphics{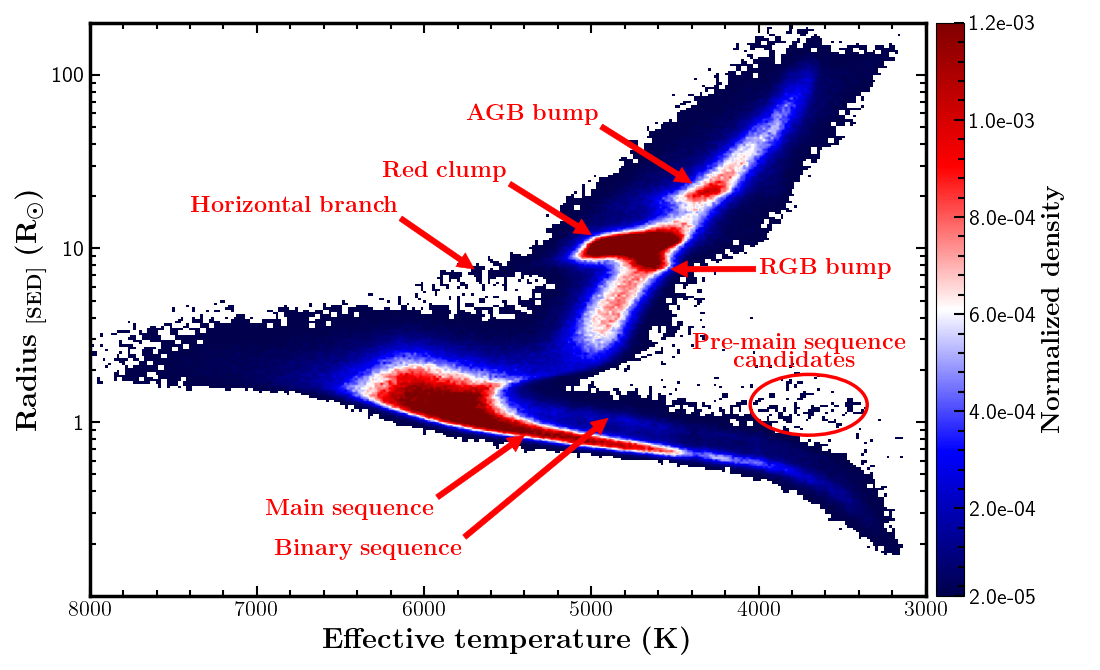}}
\caption{Radius vs \teff\ diagram, color-coded by the normalized number density. Radius estimates are derived from our SED fitting, while effective temperatures are adopted from APOGEE DR17, GALAH DR3, and RAVE DR6. Some key features are highlighted. 
}
\label{fig:hr_feature}
\end{center}
\end{figure*}

\begin{figure*}[ht!]
\begin{center}
\resizebox{0.9\textwidth}{!}{\includegraphics{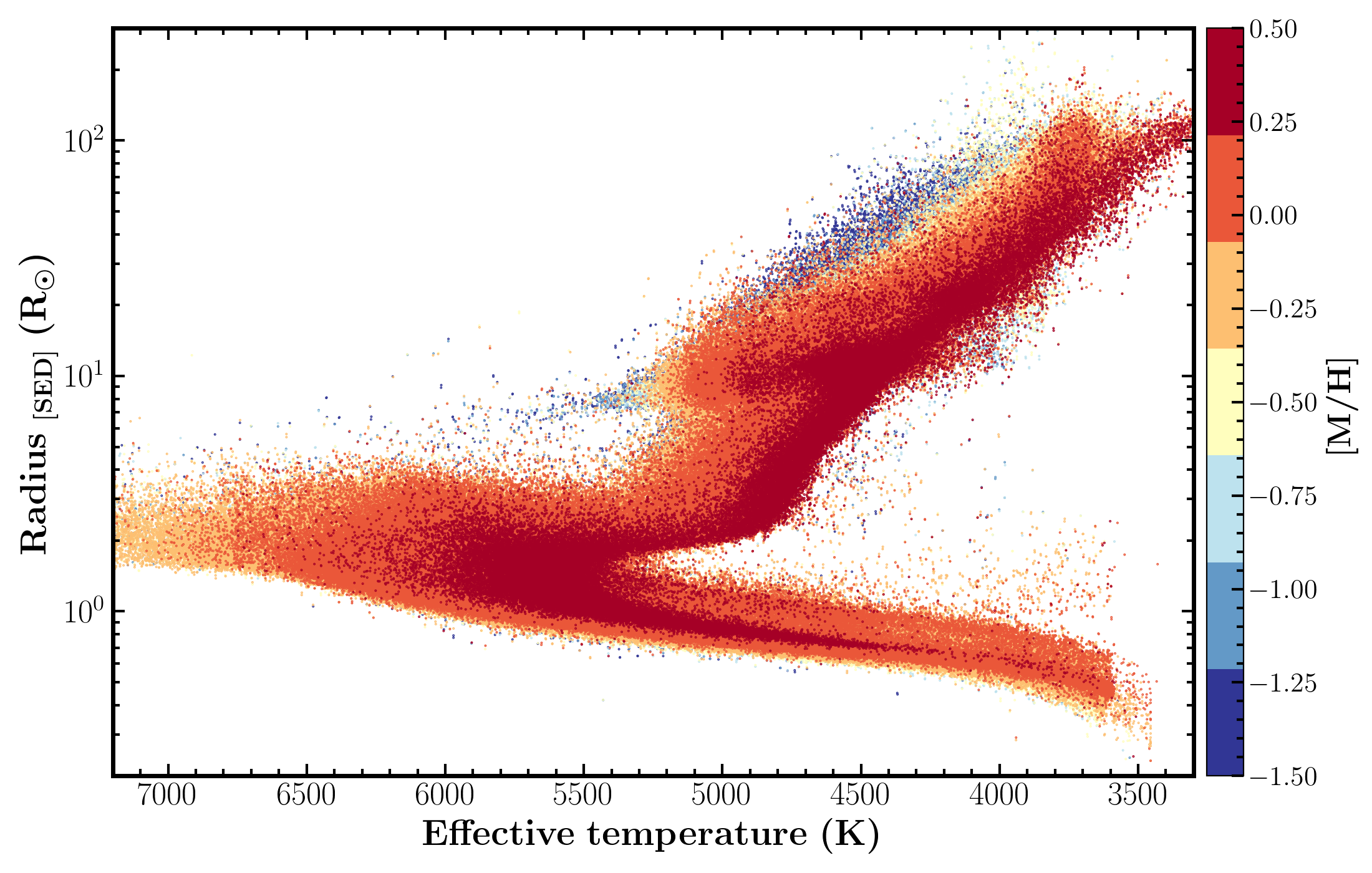}}
\caption{Similar to Figure~\ref{fig:hr_feature}, except for the color code replaced by spectroscopic metallicities from APOGEE DR17, GALAH DR3, and RAVE DR6, available for stars with \teff\ $\gtrsim$~3500 K.}
\label{fig:hr_metallicity}
\end{center}
\end{figure*}

\begin{figure*}[ht!]
\begin{center}
\resizebox{0.9\textwidth}{!}{\includegraphics{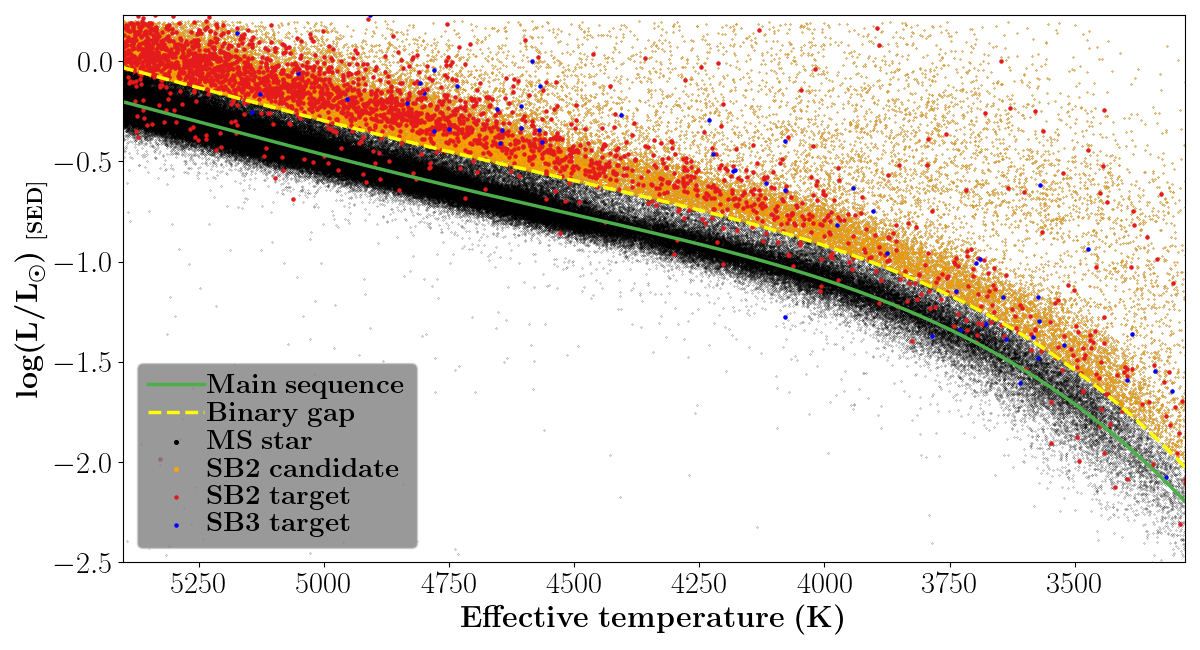}}
\caption{Hertzsprung–-Russell (H--R) diagram. The black dots mark MS stars and the orange dots denote our SB2 candidates. The red and blue points are SB2 and SB3 candidates/binaries compiled from literature,  respectively. The yellow dashed line indicates the gap between the SB2 sequence and the MS (green line, see the text).}
\label{fig:binarysequence}
\end{center}
\end{figure*}

Asteroseismology has been extensively used to test the zero points of Gaia parallaxes released in Gaia DR1, DR2, and EDR3, by comparing radius and/or parallax \citep[e.g.][]{davies2017a, huber2017a, hall2019a, khan2019a, zinn2019a, zinn2021a}. We stress that our radius scale matches the asteroseismic radius scale at a level of sub-1\%, confirming asteroseismic findings of the reduced parallax systematics in Gaia EDR3 \citep[e.g.][]{ zinn2021a} than those from the previous data releases.

\subsubsection{Luminosities}
In Figure~\ref{fig:lumivalid}, we compare the luminosities inferred from the SED fitting with those from asteroseismology \citep{serenelli2017a, pinsonneault2018a} and from the \textit{Kepler} stellar properties catalog by \citet{berger2020a}. Our luminosity estimates are consistent with the asteroseismic values, with a mean offset of \mbox{-0.004} and a dispersion of 0.042. Meanwhile, our estimates are also in good agreement with those from \citet{berger2020a}, with a mean offset of 0.008 and a dispersion of 0.053. We note that our luminosity scale is well in line with the literature studies.

\subsubsection{Extinction}
Figure~\ref{fig:Avvalid} shows a comparison between the extinctions obtained from our SED fitting, and those from the APOKASC2 catalog by \citet{pinsonneault2018a}. The comparison yields a mean offset of -0.039 and a standard deviation of 0.050.  The lower panel of Figure~\ref{fig:Avvalid} shows that the extinction residuals tend to get smaller with increasing extinction. We remind that their seismic extinctions were derived by comparing apparent magnitudes with absolute magnitudes, that were calculated from seismic luminosities and bolometric corrections. Since our luminosities are consistent with those from APOKASC2, any uncertainty in the zero points of the seismic scaling relations, if at all, should not contribute to this systematic extinction offset. The different extinction laws used in both studies should not lead to this offset either, as discussed in Section~\ref{fittingmethod}.

It is quite likely, therefore, that the main reason behind this discrepancy is related to the bolometric correction scale adopted by \citet{pinsonneault2018a}. We remind the reader that to estimate stellar parameters (e.g. luminosity) from apparent magnitudes and Gaia parallaxes, both the direct method (more specifically, bolometric corrections) and the SED-fitting method depend on the choice of spectral libraries and photometric systematic parameters (e.g. zero points of the flux densities and filter transmission curves). Any difference in these input data can lead to this small but statistically significant $A_V$ offset. We note that \citet{pinsonneault2018a} used ATLAS9, whereas we used MARCS for this sample of red giants. Furthermore, a bolometric correction is tied to a chosen reference value for absolute solar magnitude, which should be appreciated as a definition, rather than a measurement \citep{casagrande2014a}, and can thus be subject to systematic difference in literature. For example, \citet{girardi2002a} used $M_{\rm bol, \odot}$ = 4.77 mag, \citet{casagrande2014a} used $M_{\rm bol, \odot}$ = 4.75 mag, while the IAU 2015 Resolution B2 recommended $M_{\rm bol, \odot}$~=~4.74~mag. These values translate to differences in bolometric flux of up to 1.2\%.

We also compared our extinction scale with that from \citet{berger2020a}, yielding an offset of 0.02 with a scatter of 0.108 (figure not shown here). This $A_V$ offset is smaller than the offset found with respect to \citet{pinsonneault2018a}. We also note that the extinction difference between our SED fitting and \citet{berger2020a} is a stable function of extinction. We recall that the \citet{berger2020a} extinction estimates were taken from the 3D \texttt{Bayestar} dust map \citep{green2019a}. Indeed, our extinction scale is well consistent with that of \citet{green2019a} (see Sect. \ref{avscale}), in line with the comparison result 
presented here.

\section{Results}
We now present our radius and extinction estimates for 1.5 million APOGEE, GALAH, and RAVE stars.
\subsection{\teff--Radius diagram}

Figure~\ref{fig:hr_feature} shows our radius estimates in a \teff--radius diagram, color-coded by the normalized number density. Several evolutionary features stand out. In giants, we observe the red clump and its extension toward the horizontal branch up to \teff~$\simeq$~6000~K. Below the lower envelope of the red clump is the presence of the RGB bump. We recall that the RGB bump is the result of an accumulation of stars, due to a temporal decrease and subsequent increase of luminosity along the RGB.  This phenomenon is linked to the approach of the hydrogen-burning shell to the composition discontinuity left behind by the first dredge-up \citep[][and references therein]{hekker2020a}.

A morphologically similar evolutionary phase is the asymptotic-giant-branch (AGB) bump. At the beginning of the AGB phase, the He-exhausted core contracts and heats up, and the H-rich envelope expands and cools so effectively that the H-burning shell that lies above the He-burnig shell extinguishes. The expansion of the envelope is eventually halted by its own cooling and it re-contracts. The luminosity then decreases and the matter at the base of the convective envelope heats up. When the H-burning shell reignites, the luminosity increases again.  The decrease and subsequent increase of the luminosity produce a bump along the AGB \citep{ferraro1992a, gallart1998a}. A clear low-density gap between the red clump and the AGB bump is seen in Figure~\ref{fig:hr_feature}. \citet{dreau2021a} found that at the very early phase of the AGB evolution, the star evolves faster before reaching the AGB bump, leading to a lower chance to observe stars in this gap.

Another prominent feature is the binary sequence, which runs slightly above and almost parallel to the MS. It is composed of photometrically unresolved binaries. In addition, a number of APOGEE stars are present above the MS and the binary sequence. We find that these stars also stand out in the APOGEE \mbox{\teff--\logg} diagram, and are linked to rotational variables, as recognized by \citet[][]{jayasinghe2021a} with ASAS-SN light curves (see their figure~7). We find that these stars exhibit high infrared excess (traced with 2MASS K$_{S}$ and ALLWISE W3 magnitudes, \citealt{yu2021a}), and hence are pre-MS star candidates.

Our radius estimates reveal expected metallicity effects on stellar evolution, shown in Figure~\ref{fig:hr_metallicity}. We observe a metallicity gradient, particularly in red giants, where metal-poor stars have higher temperatures at a given radius than metal-rich stars do. Meanwhile, the red clump and its extension toward the horizontal branch are also visible, as is the metallicity gradient in this population.

\begin{figure*}[t]
\begin{center}
\resizebox{0.9\textwidth}{!}{\includegraphics{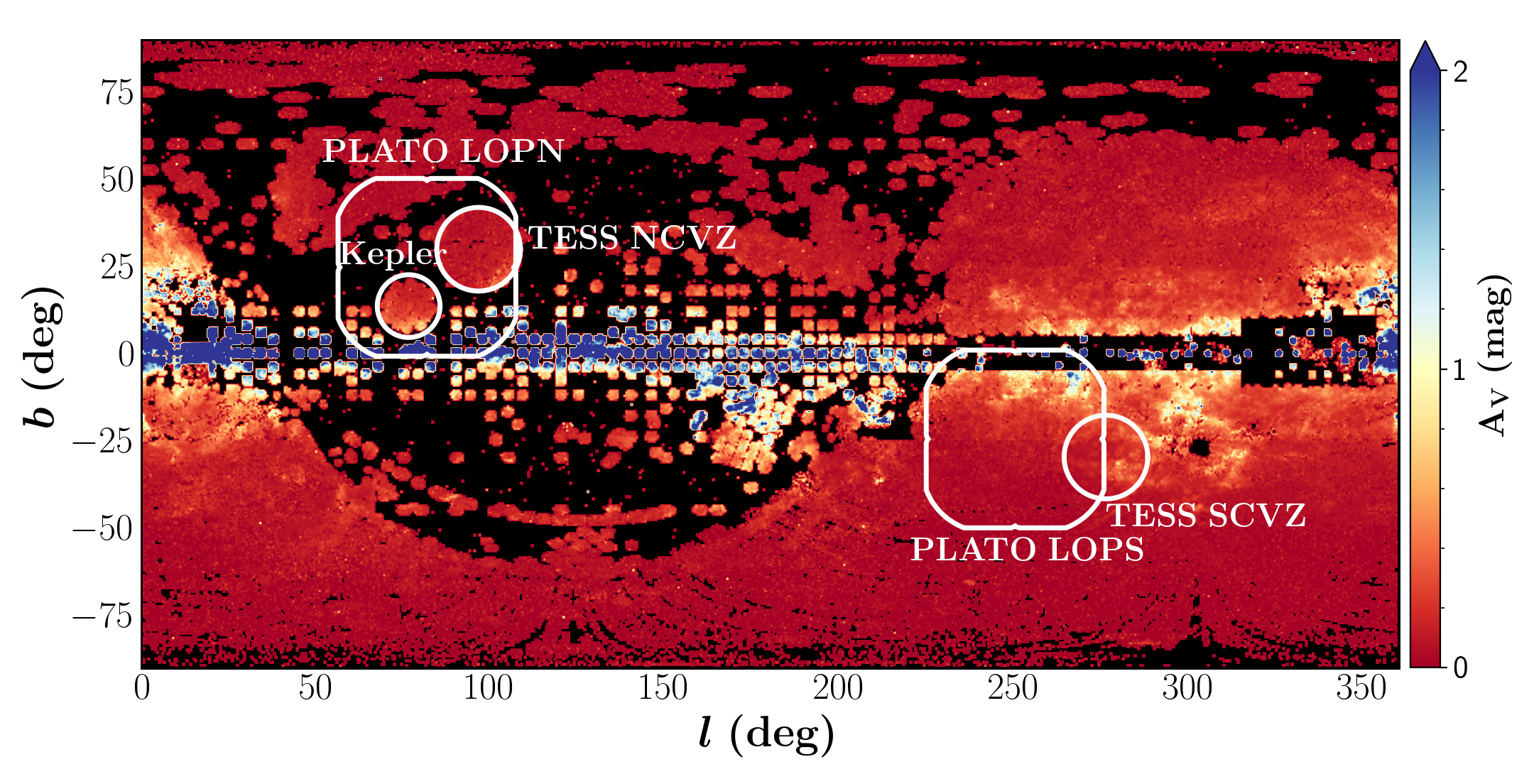}}
\resizebox{0.9\textwidth}{!}{\includegraphics{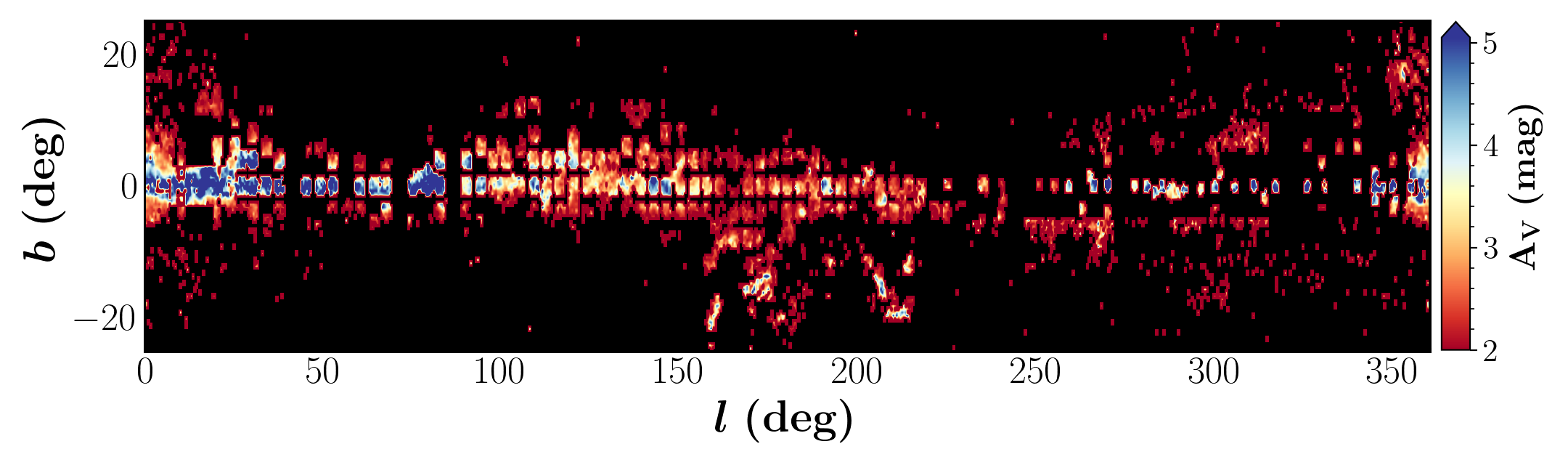}}\\
\caption{\textbf{Top:} extinction map in Galactic coordinates, traced by the footprint of APOGEE DR17, GALAH DR3, and RAVE DR6. The \textit{Kepler} field, TESS Northern and Southern Continuous Viewing Zones (NCVZ, SCVZ), and PLATO North and South Long-duration Observation Phase fields (LOPN, LOPS) are schematically highlighted. The sky region that has not been observed by the three spectroscopic surveys is shown in black. \textbf{Bottom:} Similar to the top panel, except for highlighting the galactic plane using the stars with $A_V>2$ mag (note the different $A_V$ scales of the colorbars).}
\label{fig:dustmap}
\end{center}
\end{figure*}

\begin{figure*}
\begin{center}
\resizebox{0.9\textwidth}{!}{\includegraphics{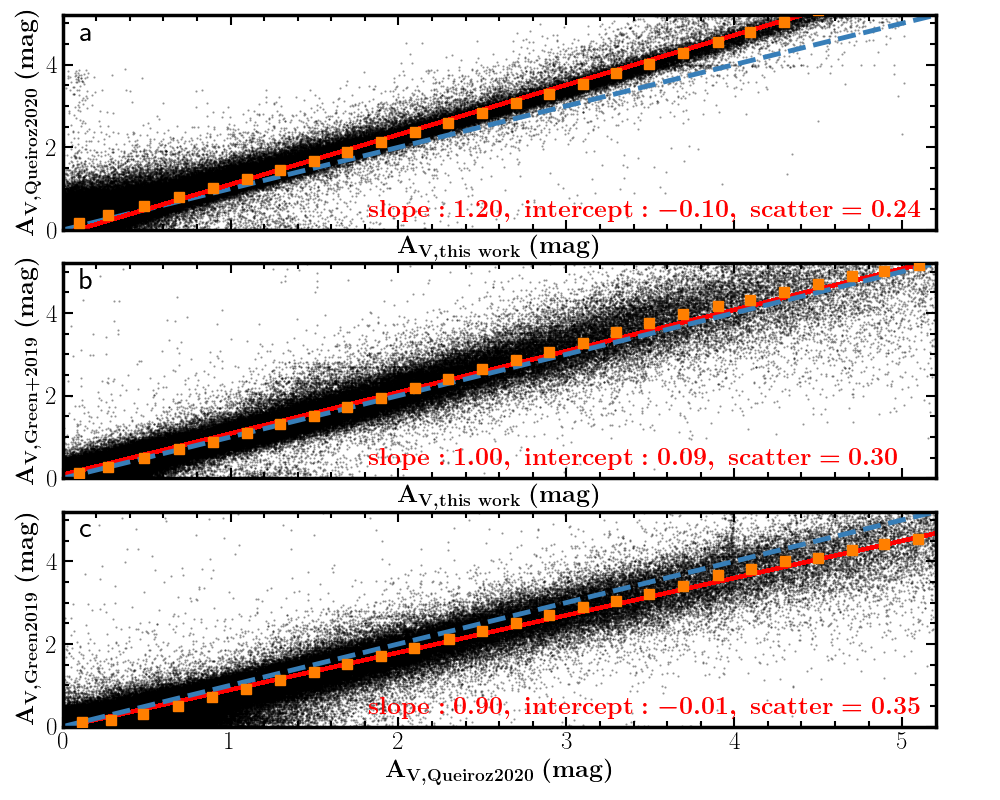}}
\caption{Comparisons between extinctions derived from the SED fitting and those from \citet{queiroz2020a} and \citet{green2019a}. In each panel, the orange diamonds show the mean values in each bin, and the red line is a fitted linear model to the orange diamonds, with its slope and intercept values annotated. The scatter value is the dispersion of the $A_V$ measurements with respect to the fitted linear model. The blue dashed line indicates the one-to-one correspondence.}
\label{fig:avcompliter}
\end{center}
\end{figure*}

Our SED-fitting method in combination with Gaia parallaxes are efficient for detecting double-lined (SB2) spectroscopic binary candidates. Figure~\ref{fig:binarysequence} shows a clear binary sequence (orange dots) located above the MS (black dots). To separate this binary sequence, first, we identified  MS stars by binning \teff\ and locating the MS star group from the bimodal distribution of luminosity in each \teff\ bin of 100~K. Then, we fitted a 4th-order polynomial to all these MS stars, shown in the green solid line\footnote{The optimized polynomial coefficients from high to low orders are -1.186E-13,  2.341E-9, -1.725E-5,  5.691E-2, and -7.216E1.}. Next, with this polynomial, we predicted luminosities given \teff\,  and subtracted these from the measured luminosities. The luminosity residuals show a bimodal distribution, where the valley is at $\delta \rm{log} (L/\rm{L_{\odot}})= 0.17$ dex. Finally, we defined the stars shown in Figure~\ref{fig:binarysequence} as SB2 candidates if their luminosities are higher than their predicted luminosities by $>$~0.17 dex.

This binary sequence well overlaps with the SB2 stars/candidates previously found with APOGEE, GALAH, RAVE, and Gaia $RVS$ spectra \citep{matijevic2010a, traven2020a, kounkel2021a, gaia-collaboration2022a}. We reported a detection of 43,054 SB2 candidates/binaries, among which 36,854 are new SB2 candidates and can be selected from Table~\ref{tab:catalog} by requiring \texttt{BinarySource} = \texttt{SED}. We recommend spectroscopic follow-up observations of these candidates to confirm the classifications and to characterize their orbital properties.

\begin{figure*}
\begin{center}
\resizebox{\textwidth}{!}{\includegraphics{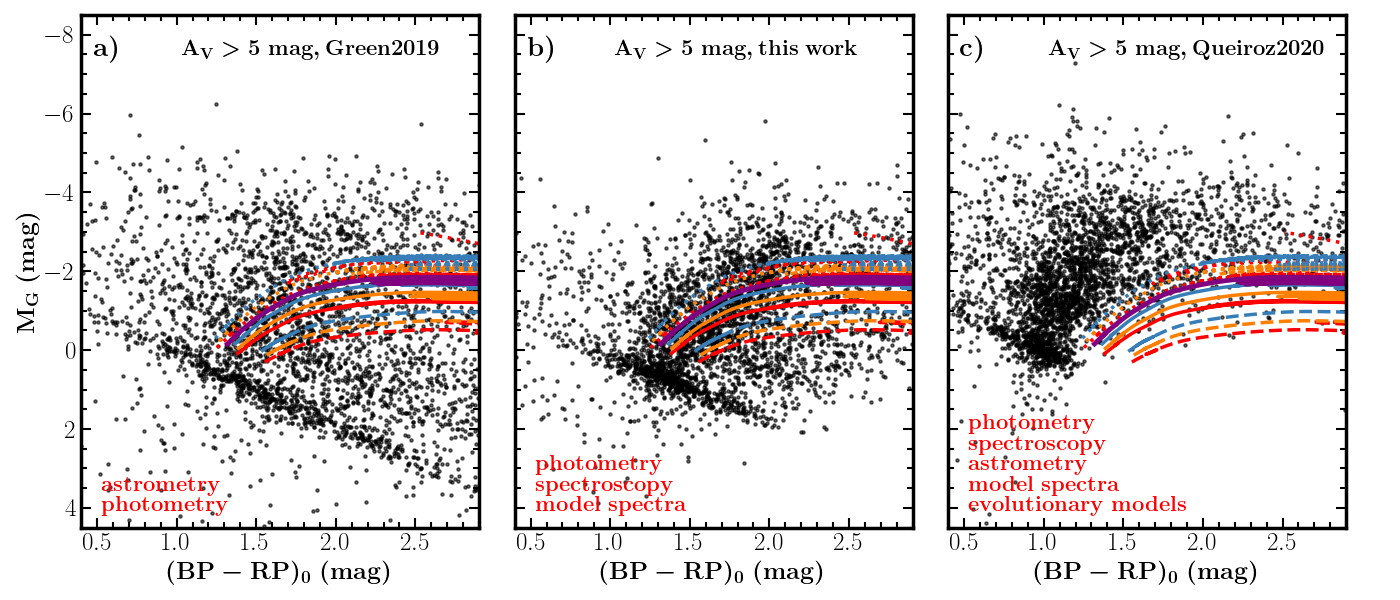}}
\caption{Dereddened Gaia color-magnitude diagram of high-extinction high-luminosity red giants ($A_V>5$ mag), using $A_V$ from \citet{green2019a} (a), this work (b), and \citet{queiroz2020a} (c). The red text indicates the data and models used for deriving $A_V$ in each work. Overplotted are MIST evolutionary tracks of the AGB phase of different reasonable masses and metallicities: dashed, solid, and dotted lines for [M/H] = 0.3, 0, -0.3, respectively, and red, orange, and blue lines for $M$ = 0.8, 1.0, 1.2\msun, respectively. The purple line corresponds to a stellar track of $M$ = 1.0\msun\ and [M/H] = 0 that has representative properties of the sample shown here. }
\label{fig:avcomphighav}
\end{center}
\end{figure*}

\subsection{All-Sky Dust Map}\label{avscale}
Figure~\ref{fig:dustmap} shows our extinction map, traced by the footprint of the three spectroscopic surveys, covering low- (RAVE), intermediate- (GALAH), and high-extinction (APOGEE) sky regions. As expected, we observe high extinctions along the Galactic plane, and lower extinctions toward higher latitudes. Differential extinction is visible in the regions dedicated for asteroseismic analyses and exoplanet studies, e.g., the \textit{Kepler} field \citep{borucki2010a}, TESS Northern and Southern Continuous Viewing Zones (NCVZ, SCVZ, \citealt{ricker2015a}), and PLATO North and South Long-duration Observation Phase fields (LOPN, LOPS, \citealt{nascimbeni2022a}). 

Next, we analyzed the accuracy and precision of our extinction estimates by comparing them with the \mbox{large-volume} extinction studies of the \texttt{StarHorse} catalog \citep{queiroz2020a}, and the 3D \texttt{Bayestar} dust map \citep{green2019a}. The \texttt{StarHorse} extinctions were obtained through a grid-based modelling approach by combining spectroscopic data from APOGEE DR16, GALAH DR2, and RAVE DR6 surveys, parallaxes from Gaia DR2, and photometry from PanSTARRS, 2MASS, and AllWISE \citep{queiroz2020a}. The \texttt{Bayestar} reddenings were acquired through a Bayesian, \mbox{model-independent} scheme by integrating Gaia parallaxes from Gaia DR2 and photometry from Pan-STARRS and 2MASS \citep{green2019a}. Note that we used $R(V)$ = 3.1, when converting \texttt{Bayestar} \textit{E(B-V)} to $A_V$ via $A_V=R(V)E(B-V)$, for consistency (see Sect. \ref{fittingmethod}). This is equivalent to comparing our \textit{E(B-V)} estimates with those from \citet{green2019a}.

In Figure~\ref{fig:avcompliter}a, we compare our $A_V$ estimates that are independent of stellar evolutionary models with those from \citet{queiroz2020a}, yielding a tight correlation with a scatter of 0.24. However, a significant scale difference is visible, with a slope of 1.20 and an intercept of -0.10. It is important to recall that \citet{queiroz2020a} used different data releases of spectroscopic data sets (APOGEE DR16, GALAH DR2, RAVE DR6) than ours (APOGEE DR17, GALAH DR3, RAVE DR6). We then repeated our fitting analysis with their input spectroscopic data, and found that the significant $A_V$ scale difference remains. Specifically, the $A_V$ difference caused by the different spectroscopic data releases is 0.01$\pm$0.13~mag, without any systematic trend. Since our SED fitting analysis is more sensitive to \teff\ than \logg\ and metallicity, we conclude that the different \teff\ scales, which is $18\pm93$ K, cannot lead to this significantly different $A_V$ scales shown in Figure~\ref{fig:avcompliter}a. 

In Figure~\ref{fig:avcompliter}b, we then compare our $A_V$ estimates with those from \citet{green2019a}, yielding a larger dispersion of 0.30 but with a good scale consistency (slope = 1.00, intercept=0.09). This dispersion is smaller when comparing the extinctions from \citet{queiroz2020a} and \citet{green2019a} ($\sigma$ = 0.35), as shown in Figure~\ref{fig:avcompliter}c. This suggests that our extinction estimates are more precise than those from \citet{queiroz2020a}. Our choice of using 32 bandpasses, which are more than those used in \citet{queiroz2020a}, is likely to be responsible for this.

\begin{deluxetable}{rrrrrrr}[t]
\tablecaption{Global extinctions of 184 Gaia open clusters\label{tab:opencluster}}
\tablewidth{0pt}
\tablehead{
\colhead{Cluster} & \colhead{$l$} & \colhead{$b$} & \colhead{$A_{\rm {V,NN}}$} &
\colhead{$A_{\rm V}$} & \colhead{$\sigma _{\rm {A_V}}$} & \colhead{\textit{N}}\\
\colhead{} & \colhead{deg} & \colhead{deg} &\colhead{mag}  & \colhead{mag} & \colhead{mag} & \colhead{}
}
\startdata
     ... & ... & ... & ... & ... & ... & ...\\ 
     NGC 6819 &  73.982 &   8.481 & 0.40 &  0.53 &  0.16 &       49 \\
     NGC 1817 & 186.193 & -13.032 & 0.59 &  0.64 &  0.15 &       49 \\
     Blanco 1 &  15.090 & -79.086 & 0.01 & -0.01 &  0.13 &       51 \\
     NGC 2204 & 226.016 & -16.114 & 0.01 &  0.27 &  0.17 &       59 \\
      NGC 188 & 122.837 &  22.373 & 0.21 &  0.29 &  0.17 &       60 \\
     NGC 2244 & 206.361 &  -2.026 & 1.46 &  1.34 &  0.19 &       60 \\
     NGC 2158 & 186.635 &   1.788 & 1.44 &  1.46 &  0.18 &       67 \\
 Ruprecht 147 &  20.930 & -12.760 & 0.06 &  0.30 &  0.16 &       68 \\
     NGC 6791 &  69.964 &  10.906 & 0.70 &  0.41 &  0.19 &       77 \\
   Melotte 20 & 147.357 &  -6.404 & 0.30 &  0.23 &  0.15 &       81 \\
     NGC 7789 & 115.527 &  -5.366 & 0.83 &  0.87 &  0.15 &       88 \\
      NGC 752 & 136.959 & -23.289 & 0.07 &  0.12 &  0.16 &      110 \\
   Melotte 25 & 179.767 & -21.164 & 0.00 &  0.11 &  0.15 &      118 \\
Collinder 261 & 301.696 &  -5.537 & 0.81 &  0.90 &  0.14 &      216 \\
     NGC 2168 & 186.609 &   2.230 & 0.46 &  0.61 &  0.15 &      225 \\
     NGC 2632 & 205.952 &  32.428 & 0.00 &  0.11 &  0.15 &      228 \\
   Melotte 22 & 166.462 & -23.614 & 0.18 &  0.17 &  0.15 &      269 \\
     NGC 2682 & 215.691 &  31.921 & 0.07 &  0.11 &  0.16 &      350 \\     
\enddata
\tablecomments{The 2nd and 3rd columns are the Galactic latitude and longitude of the center of each cluster, respectively, and the 4th column is the extinction estimate, all taken from \citet{cantat-gaudin2020a}. The 5th, 6th, and 7th columns are extinction estimate, its uncertainty, and number of members, all from this work. See the text for the definition of the extinction uncertainty. Only those clusters that have at least 49 members observed by APOGEE, GALAH, or RAVE are shown here. The entire table for 184 Gaia open clusters with at least 3 members each is available online.}
\end{deluxetable}

In Figure~\ref{fig:avcompliter}a, we observe a number of stars with low extinctions from our work and high values determined by \citet{queiroz2020a}, namely the stars located above the red line   ($A_V \lesssim$ 0.5 mag, lower left corner). These stars are present below the red line in the lower left corner of Figure~\ref{fig:avcompliter}c. Given that these stars are not outliers in Figure~\ref{fig:avcompliter}b, this would suggest that for these stars the extinctions of \citet{queiroz2020a} are overestimated.

In addition to the feature in the low extinction regime, a larger scatter is visible in the high extinction range, as shown in Figs.~\ref{fig:avcompliter}b and ~\ref{fig:avcompliter}c.
Now, we focus on the sample of stars with high extinctions of $A_V>5$ mag to understand the precision of each extinction scale. These stars are shown in dereddened color-magnitude diagrams in Figure~\ref{fig:avcomphighav}. First, we see that the AGB population is the most scattered if the extinctions of \citet{green2019a} are used (Figure~\ref{fig:avcomphighav}a); the population is less scattered when our extinctions are used (Figure~\ref{fig:avcomphighav}b), and is the least scattered when the extinctions of \citet{queiroz2020a} are used (Figure~\ref{fig:avcomphighav}c). The same conclusion applies to the scatter (elongation) of the red clump. The different $A_V$ precisions rely on the amount of observational data used (photometry, spectroscopy, and/or astrometry), as well as on the extent of stellar models involved (see the text legend in Figure~\ref{fig:avcomphighav}). Overplotted are the MIST evolutionary models of the AGB phase with various reasonable masses ($M$ = 0.8, 1.0, 1.2\msun) and metallicities ([M/H] = 0.3, 0, -0.3) \citep{dotter2016a, choi2016a}. We can see that the models better match the observations if our extinctions, or those from \citet{green2019a} are used.

\begin{figure}
\begin{center}
\resizebox{\columnwidth}{!}{\includegraphics{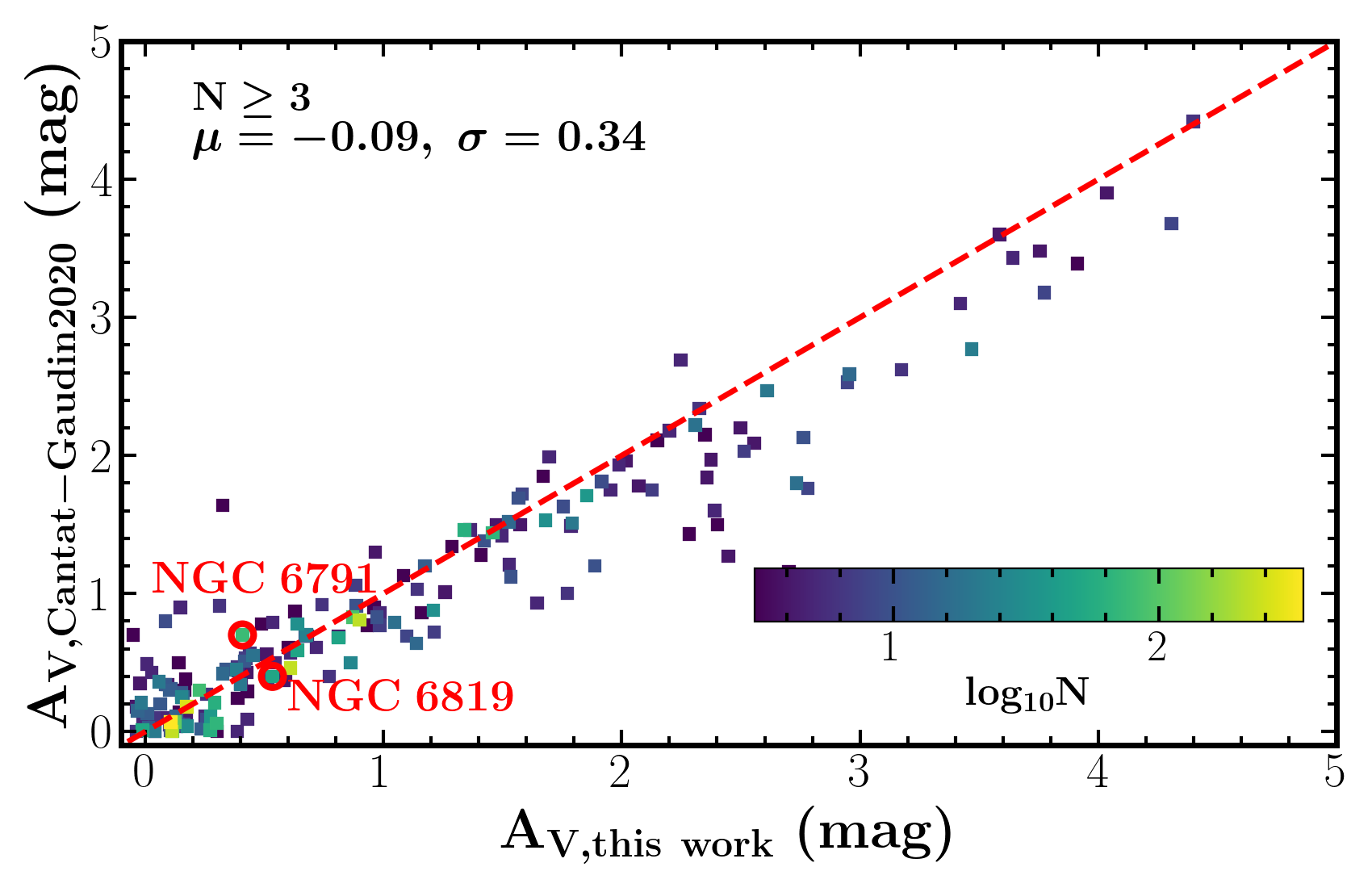}}
\caption{Comparison of extinctions of 184 Gaia open clusters that have been investigated by \citet{cantat-gaudin2020a} and that have at least 3 members each in our sample. The color code indicates the number of the cluster members of each cluster. Two open clusters,  NGC 6791 and NGC 6819, that are subject to extensive asteroseismic analysis are highlighted in red open circles. The red dashed line shows the one-to-one relation. The mean difference (-0.09) and its standard deviation (0.34) are indicated, where the literature extinction scale is on average less than ours. }
\label{fig:clusterAvComp}
\end{center}
\end{figure}

\begin{figure*}
\begin{center}
\resizebox{1.05\columnwidth}{!}{\includegraphics{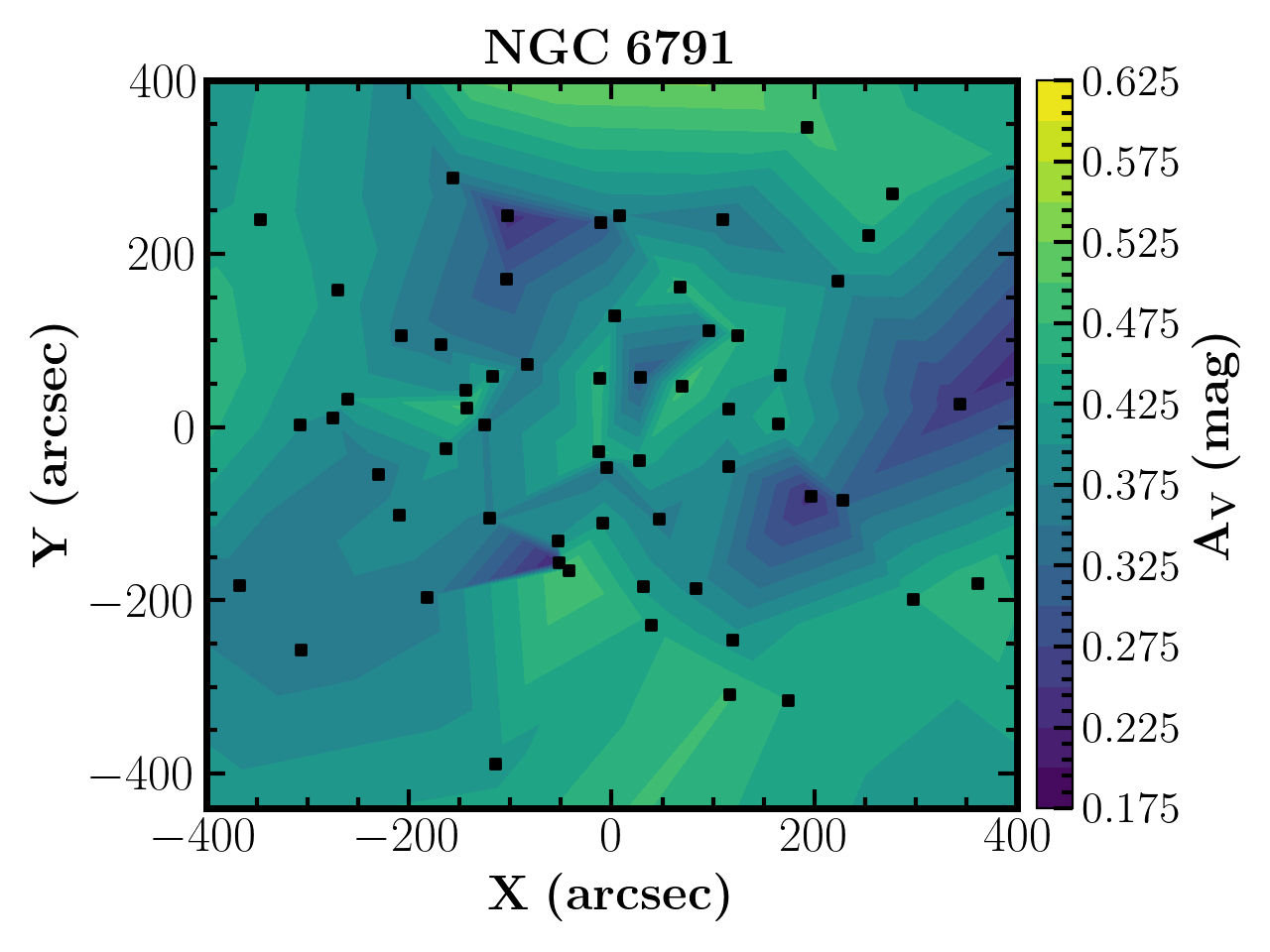}}
\resizebox{1.05\columnwidth}{!}{\includegraphics{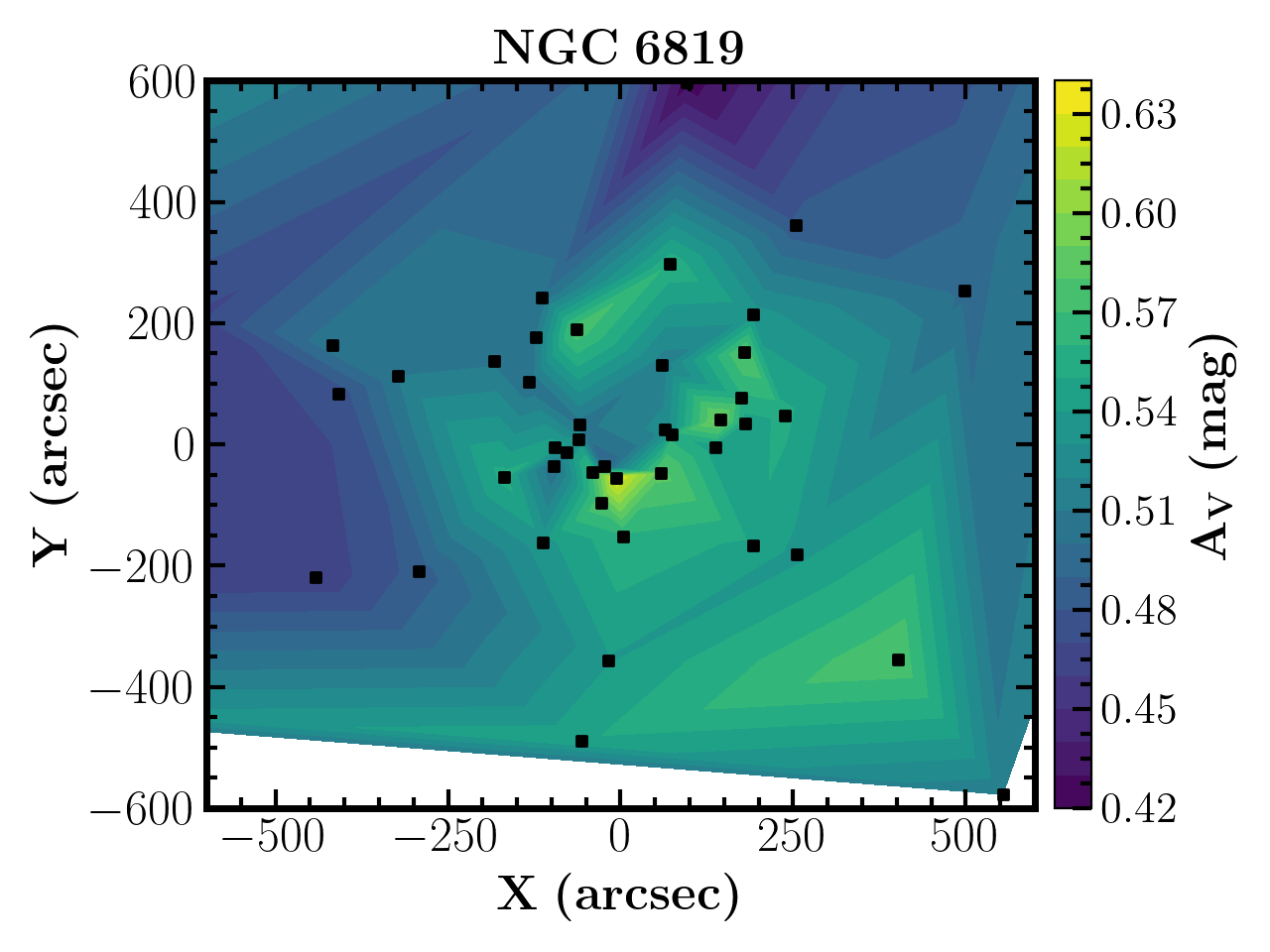}}
\caption{Spatial distributions of the extinctions of NGC 6791 with respect to the cluster center (ra, dec)=(295.32283, 40.19639), and of NGC~6819 with respect to the cluster center (ra, dec)=(295.32283, 40.19639). The black points indicate the cluster star members \citep{cantat-gaudin2020a} in our sample that are used to probe differential extinction as shown in the filled contours. The margin at the bottom of the right panel is not covered by any stars. For comparison, our field windows shown here are chosen to be the same as \citet{brogaard2012a} for NGC 6791 and as \citet{platais2013a} for NGC 6819.}
\label{fig:diffAv}
\end{center}
\end{figure*}

In summary, our extinction scale is consistent with \citet{green2019a} at $<$1\% level, and deviates from \citet{queiroz2020a} by $\sim$20\%. Our extinction scale and \citet{green2019a} better matches the MIST evolutionary models than \citet{queiroz2020a}. Globally, our extinctions have the highest precision (see the scatter values annotated in Figure~\ref{fig:avcompliter}), probably due to the use of a higher number of photometric bandpasses. The extinction values from our work and \citet{green2019a} are more consistent in low-extinction regimes ($A_V \lesssim$ 0.5 mag). Based on the analysis above, we keep our extinction scale as it is, i.e., without a calibration onto a reference scale. 

\subsection{Extinctions of Gaia Open Clusters} \label{openclusterav}
After fixing our extinction scale, we then used the extinctions of individual stars to determine the global extinctions of 184 Gaia open clusters, with the membership classifications from  \citet{cantat-gaudin2020a}. Each of these clusters has at least 3 members in our sample. We adopted the median extinction of the stars in each cluster to estimate the global cluster extinction. We approximated its uncertainty by summing up in quadrature the median formal uncertainty and a statistical uncertainty $\sigma/\sqrt{N}$, where $\sigma$ is the standard deviation of the extinction estimates of cluster members and $N$ is the number of the stars in the cluster. Table~\ref{tab:opencluster} lists the extinctions and their uncertainties for these 184 Gaia open clusters. 

Figure~\ref{fig:clusterAvComp} shows a comparison of the extinctions for 184 open clusters derived from our work with those from \citet{cantat-gaudin2020a}. \citet{cantat-gaudin2020a} estimated their average cluster extinctions by training an artificial neural network with Gaia photometry (i.e., color-magnitude diagrams) and Gaia parallaxes. We observe a good consistency in Figure~\ref{fig:clusterAvComp}, with a mean difference of -0.09 mag and a standard deviation of 0.34 mag, in the sense that the \citet{cantat-gaudin2020a} extinction scale is on average slightly lower than ours.

Figure~\ref{fig:clusterAvComp} also shows that in the high extinction regime ($A_{V}\gtrsim2.5$ mag) the $A_{V}$ estimates from \citet{cantat-gaudin2020a} are globally smaller than ours. This is consistent with their findings when comparing their predicted $A_{V}$ estimates with those from their training sample and with the independent literature study in \citet[][see their figure 5]{kharchenko2013a}. The reason for this $A_{V}$ underestimation, is that, there are few open clusters with $A_{V}\gtrsim2.5$ mag in their training set. It is well known that statistical inference on small samples leads to inferior estimates in training and consequently to reduced performances in prediction.

\begin{figure*}[ht!]
\begin{center}
\resizebox{\textwidth}{!}{\includegraphics{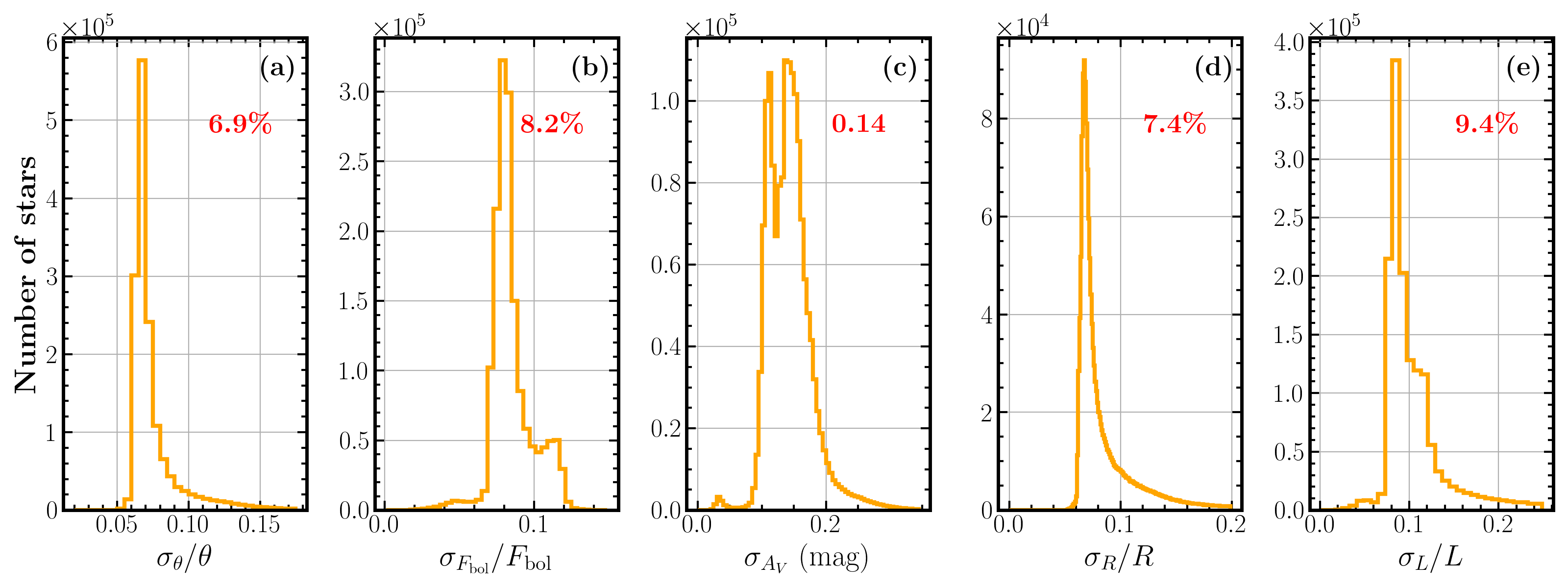}}
\caption{Uncertainty distributions of angular radius, bolometric flux, extinction, radius, and luminosity (from left to right). The numbers shown in red are the median uncertainties.}
\label{fig:error}
\end{center}
\end{figure*}

Next, we discuss two open clusters, NGC 6791 and NGC 6819, that have gained tremendous importance and have been subject to extensive asteroseismic analyses due to the availability of exquisite \textit{Kepler} data. For NGC 6791, our analysis yields \mbox{\textit{E(B-V)} = 0.13 $\pm$ 0.06 mag} (or $A_V$=\mbox{0.41 $\pm$ 0.19 mag}, adopting $R(V)$ = 3.1), consistent with the literature value of  \mbox{\textit{E(B-V)} = 0.10--0.16} \citep[][and references therein]{an2015a} and close to \mbox{\textit{E(B-V)} = 0.16 $\pm$ 0.025 mag}, as obtained from asteroseismic analysis by \citet{wu2014a} and \citet{brogaard2021a}. We note that \citet{brogaard2021a} also  reported a second value, \mbox{\textit{E(B-V)} = 0.13 mag} supported by their analysis, in line with ours. For NGC 6819, we obtained a reddening value of \mbox{\textit{E(B-V)} = 0.17$ \pm$ 0.05 mag} (or \mbox{$A_V$ = 0.53 $\pm$ 0.16 mag}). This is consistent with the asteroseismic value evaluated by \citet[][]{handberg2017a}, \mbox{\textit{E(B-V)} = 0.15 mag}, as well as the results in \citet{bragaglia2001a}, \mbox{\textit{E(B-V)} = 0.142 $\pm$ 0.044 mag}, and \citet{rosvick1998a}, \mbox{\textit{E(B-V)} = 0.16 mag}.

Our extinction estimates of the members of NGC~6791 and NGC~6819 allow insights into their potential differential reddening, a feature that is usually identified by the spread of the upper MS in \mbox{color--magnitude} diagrams. Indeed, \citet{brogaard2012a} found evidence in favor of differential reddening in NGC 6791, but warned that systematic effects from instrument photometry and/or the reduction procedure may also be present in their findings. We note that their reported differential reddening is \mbox{$\Delta$E(B-V) = $\pm$0.04 mag}, which is at the same level as their color precision. As shown in Figure~\ref{fig:diffAv}, our extinction estimates reproduce the major differential extinction features seen in \citet{brogaard2012a}. For example, for NGC 6791, we also find low extinction around (X,Y) = (200, 100). Meanwhile, \citet{platais2013a} found a maximum differential reddening of \mbox{$\Delta$\textit{E(B-V)} = 0.06 mag} in NGC 6819. A key feature revealed therein is the presence of a local high extinction region, expected in the direction from the cluster center toward the bottom right corner of the right panel of Figure~\ref{fig:diffAv}. Indeed, our results support the finding by \citet{platais2013a}. While our analysis of differential reddening is limited by the spatial resolution due to a relatively small sample size, the high precision of extinction measurements enables us to confirm the existence of the differential extinction in both NGC~6791 \mbox{($A_V=0.18$~to~$0.63$ mag)} and NGC~6819 ($A_V=0.42$~to~$0.63$~mag).

\begin{deluxetable*}{lll}[ht!]
\tablecaption{SED-fitting-based stellar parameters of $\sim$1.5 million stars\label{tab:catalog}}
\tablewidth{0pt}
\tablehead{\colhead{Label} & \colhead{Units} & \colhead{Description}}
\startdata
\texttt{starID} & & \texttt{APOGEE\_ID} for APOGEE, \texttt{sobject\_id} for GALAG, and \texttt{raveid} for RAVE \\
\texttt{Gaia\_DR3\_ID} & & Gaia DR3 or EDR3 \texttt{source\_id}\\
\texttt{teff} & Kelvin & calibrated \teff \\
\texttt{tefferr} & Kelvin & re-scaled \teff\ uncertainties, including a 2.4\% error floor determined with interferometry. \\
\texttt{logg} & dex & calibrated \logg \\
\texttt{loggerr} & dex & re-scaled \logg\ uncertainties \\
\texttt{feh} & dex & calibrated [Fe/H] \\
\texttt{feherr} & dex & re-scaled [Fe/H] uncertainties \\
\texttt{d} & pc & \citet{bailer-jones2021a} distances, \texttt{r\_med\_photogeo} preferred if available, otherwise \texttt{r\_med\_geo} \\
\texttt{derr} & pc & \makecell{\citet{bailer-jones2021a} distance uncertainties, $(\texttt{r\_hi\_photogeo}-\texttt{r\_lo\_photogeo})/2$ preferred \\  if available, otherwise $(\texttt{r\_hi\_geo}-\texttt{r\_lo\_geo})/2$}\\ 
\texttt{Av} & mag & extinction in V, assuming R(V)=3.1 \\
\texttt{Averr} & mag & extinction uncertainty in V\\
\texttt{Fbol} & erg/s/cm$^2$ & bolometric flux\\
\texttt{Fbolerr} & erg/s/cm$^2$ & bolometric flux uncertainty\\
\texttt{angRad} & mas & anguar radius \\
\texttt{angRaderr} & mas & anguar radius uncertaity\\
\texttt{lumi} & dex & luminosity, $\rm{log}(\it{L}/\rm{L_{\odot}})$\\
\texttt{lumierr} & dex & luminosity uncertainty\\
\texttt{radius} & $R_{\odot}$ & radius \\
\texttt{radiuserr} & $R_{\odot}$ & radius uncertainty\\
\texttt{npoint} & & number of photometric points used for the SED fitting\\
\texttt{source} & & target source, i.e., APOGEE DR17, GALAH DR3, or RAVE DR6\\
\texttt{ruwe} & & Gaia Re-normalized Unit Weight Error \\
\texttt{ra} & degree & right ascension (J2000)\\
\texttt{dec} & degree & declination (J2000) \\
\texttt{l} & degree & galatic lattitude\\
\texttt{b} & degree & galactic longitude\\
\texttt{Uniq}  & & whether unique, Y or N, priority: APOGEE $>$ GALAH $>$ RAVE\\
\texttt{BinarySource} & & NaN, SED, APOGEE, GALAH, RAVE, GAIA, or the combinations\\
\enddata
\tablecomments{Among the 1,566,810 entries in this catalog, there are 1,484,987 unique stars. The entire table is accessible online.}
\end{deluxetable*}

\section{Catalog}\label{catalog}
We present our final catalog in Table \ref{tab:catalog}. It offers homogenized atmospheric parameter estimates and their uncertainties for stars in APOGEE DR17, GALAH DR3, and RAVE DR6. It also provides derived parameter values from our SED-fitting analysis, namely, radii, luminosities, extinctions, bolometric fluxes, and angular radii, together with other datasets for user convenience.

Figure~\ref{fig:error} shows the resultant uncertainty distributions of our derived parameters. Our catalog median precision is 6.9\% for angular radius, 8.2\% for bolometric flux, 0.14 mag for extinction, 7.4\% for radius, and 9.4\% for luminosity (0.04 dex for \logl). Note that we added in quadrature a 2.4\% uncertainty floor to the random \teff\ uncertainties \citep{tayar2022a}, which affects all of the conservative uncertainty distributions shown in Figure~\ref{fig:error}. The distribution of extinction uncertainty is bimodal. This is because the majority of the APOGEE stars (95.5\%), which are globally distant, are included in the right peak, while GALAH and RAVE stars, which are typically nearby, are nearly equally distributed in both peaks.

Our catalog is subject to two caveats. First, the derived parameters for photometrically unresolved binaries could be biased by the combined photometry of binary components. For this, we added in Table \ref{tab:catalog} the column \texttt{BinarySource}, to select/reject confirmed or candidate binary systems. This  lists SB2 binary candidates found in our work (See section 4.1 and Figure 10) and previously detected SB2/SB3 systems from literature using APOGEE \citep{kounkel2021a}, GALAH \citep{traven2020a}, and RAVE \citep{matijevic2010a} spectra. For convenience, we also added Gaia DR3 binaries \citep[\texttt{non\_single\_star=1} in table \texttt{gaiadr3.gaia\_source},][]{creevey2022a} and Gaia EDR3 re-normalized unit weight error (\texttt{RUWE}) values in Table \ref{tab:catalog}. We note that while we listed Gaia binaries, some binaries, such as SB1, should not significantly bias our SED fitting, as the fluxes of the primaries are not dramatically influenced by their companions. Second, our stellar parameter estimates for B stars, from APOGEE DR 17, should be used with caution. This is because near-UV photometry is absent to constrain their SEDs.

\section{Conclusions}
We report revised stellar radii and extinctions for 1,484,498 unique stars in the low- to high-extinction fields observed by the APOGEE, GALAH, and RAVE surveys (Table~\ref{tab:catalog} for data access). Specifically, we compare SEDs predicted by the widely used MARCS and BOSZ model spectra with 32 large-volume photometric bandpasses, combining data from 9 major surveys: Gaia EDR3, 2MASS, ALLWISE, SkyMapper, Pan-STARRS, SDSS, APASS, Hipparcos, and Tycho2. Our careful compilation of the zero points and transmission curves of these photometric systems, shown in Table~\ref{tab:photoParams}, allows us to obtain high accuracy in the derived stellar parameter estimates. This work restricts the analysis to targets with available spectroscopy, as the availability of their spectroscopic \teff\ estimates allows one to lift the temperature-extinction degeneracy. 

The primary goals of this work are fourfold. (1) To design, validate, and demonstrate our publiclly accessible pipeline, SEDEX\footnote{\url{https://github.com/Jieyu126/SEDEX}}, for carrying out SED fitting using data sets from massive surveys. (2) To provide stellar parameter estimates, such as radii and luminosities, which are critical, independent observational constraints for asteroseismic modelling with \textit{Kepler}, TESS, and future PLATO data. (3) To provide a large sample of (36,854) SB2 \mbox{MS} binary candidates. (4) To obtain interstellar extinctions and use them in our future work for constructing 3D dust maps for space-borne transit survey fields, such as the \textit{Kepler} field, the TESS CVZs, and the PLATO LOPN and LOPS fields. This would be useful for deriving stellar parameters for asteroseismic analyses and exoplanet studies.

Our results are summarized below:
\begin{itemize}
    \item We homogenize cross-survey atmospheric parameters by calibrating GALAH and RAVE scales to those of APOGEE using neural networks. We also re-scale the heterogeneous atmospheric parameter uncertainties (Sect. \ref{homoparams} and Table\ref{tab:errors}).

    \item Our validation reports consistency with CHARA angular diameters (3.07\%$\pm$6.77\%), HST CALSPEC bolometric flux (3.41\%$\pm$1.99\%), and asteroseismic extinction ($-$0.039$\pm$0.050), radius (0.1\%$\pm$4.9\%), and luminosity ($-0.4$\%$\pm$4.2\%) (Sect. \ref{validation}).
    
    \item We provide extinction estimates of the stars in APOGEE DR17, GALAH DR3, and RAVE DR6 data sets. Our extinction scale is in agreement with \citet{green2019a} at  $<$1\% level, but deviates from \citet{queiroz2020a} by $\sim$20\%.
    The extinctions from our work and \citet{green2019a} produce color-magnitude diagrams that better match the MIST stellar evolutionary models.

    \item We provide extinction values for 184 Gaia open clusters, each with at least three cluster members included in our sample. We then confirm the presence of  differential extinction in NGC~6791 ($A_V=0.18$~to~$0.63$~mag) and NGC~6819 ($A_V=0.42$~to~$0.63$~mag) (Sect.~\ref{openclusterav}). We obtain a global extinction value of $A_V$=\mbox{0.41 $\pm$ 0.19} mag for NGC~6791 and of \mbox{$A_V$ = 0.53 $\pm$ 0.16 mag} for NGC~6819.
    
    \item Our catalog median precision is 6.9\% for angular radius, 8.2\% for bolometric flux, 0.14 mag for extinction, 7.4\% for radius, and 9.4\% for luminosity (Sect.~\ref{catalog}). Note that we added in quadrature a 2.4\% uncertainty floor to the \teff\ uncertainties, which affects all of the uncertainty reported here.
\end{itemize}
Gaia Data Release 3 took place in  June 2022. Among other products, this catalog provides precise spectroscopic atmospheric parameters for $\sim$5.6 million stars based on their high-resolution $RVS$ spectra. Given that this data release does not comprise extinctions based on the high-resolution spectroscopic observations, in our next paper (Yu et al. in prep), we will use the atmospheric parameters from this data release to infer precise stellar radii, and derive a 3D extinction map. This map will cover the entire sky, beyond the current footprint of the APOGEE, GALAH, and RAVE surveys. Since this map covers several transit survey fields, such as the \textit{Kepler} field, the TESS northern and southern CVZs, and the PLATO LOPN and LOPS fields, it will be useful for exoplanet studies and asteroseismic analysis. 

Future Gaia Data Releases will enable the straightforward updates of stellar radii and luminosities to achieve better precision by combining Gaia distances with angular radii and bolometric fluxes retrieved from this work. The combination of these luminosity estimates with global seismic parameters will be valuable for deriving stellar ages for Galactic archaeology (see the introduction). Our \texttt{SEDEX} pipeline can provide fundamental stellar parameters, which might be valuable to build stellar input catalogs for future missions, such as PLATO and Earth~2.0 \citep{ge2022a}.

\bigbreak
We are greatly thankful for the referee’s thorough review of the manuscript and helpful comments. We thank Nadiia Kostogryz, Ren\'e Heller, Daniel Huber, Yuan-Sen Ting, Maosheng Xiang, and Haibo Yuan for discussions. J.Y. and L.G. acknowledge support from ERC Synergy Grant WHOLE SUN 810218 and PLATO grants from the German Aerospace Center (DLR 50OO1501) and from the Max Planck Society. SK acknowledges support from the European Union's Horizon 2020 research and innovation program under grant agreement No 101004110, and from the Netherlands Organisation for Scientific Research (NOVA). This work was funded by the Deutsche Forschungsgemeinschaft (DFG, German Research Foundation) -- Project-ID 138713538 -- SFB 881 (``The Milky Way System'', subproject P02). S.H. acknowledges the ERC Consolidator Grant DipolarSound (grant agreement \# 101000296). S.L.B. and J.Y. acknowledge the Joint Research Fund in Astronomy (U2031203) under a cooperative agreement between the National Natural Science Foundation of China (NSFC) and Chinese Academy of Sciences (CAS).

\bibliography{reference}{}
\bibliographystyle{aasjournal}
\end{document}